\documentclass[12pt]{JHEP3}
\usepackage{multicol}
\usepackage[dvips]{epsfig,graphics}
\usepackage{graphicx}
 \newlength{\wth}
 \newcommand{\twographs}[2]{%
 \unitlength=1.1in
 \begin{picture}(5.8,2.6)
 \put(0,0){\epsfig{file=#1, width=\wth}}
 \put(2.7,0){\epsfig{file=#2, width=\wth}}
 \put(0,2.1){(a)}
 \put(2.7,2.1){(b)}
 \end{picture}
}

\newcommand{\twographsg}[2]{%
 \unitlength=1.1in
 \begin{picture}(6.,2.6)
 \put(-0.2,0){\epsfig{file=#1, width=\wth}}
 \put(3.,0){\epsfig{file=#2, width=\wth}}
 \put(0,2.1){(a)}
 \put(3.0,2.1){(b)}
 \end{picture}
}

\newcommand{\fourgraphs}[4]{%
\unitlength=1in
\begin{picture}(5.8,5.2)
\put(0,0){\epsfig{file=#3.eps, width=\wth}}
\put(2.7,0){\epsfig{file=#4.eps, width=\wth}}
\put(0,2.6){\epsfig{file=#1.eps, width=\wth}}
\put(2.7,2.6){\epsfig{file=#2.eps, width=\wth}}
\put(0.1,2.3){(c)}
\put(0.1,4.9){(a)}
\put(3.0,2.3){(d)}
\put(3.0,4.9){(b)}
\end{picture}}

\title{Requirements on  collider data to match the
precision of WMAP on  supersymmetric dark matter.}

\author{B.C. Allanach\footnote{address after 1st Oct: DAMTP, CMS, Wilberforce Road,
Cambridge, CB3 OWA, UK.}, G.~B\'elanger, F.~Boudjema \\
LAPTH, 9 chemin Bellevue, BP110, F-74941 Annecy-le-Vieux, C\'edex, France.\\
}
\author{A. Pukhov\\
Skobeltsyn Institute of Nuclear Physics, Moscow State University,
Moscow, Russia.}

\abstract{If future colliders discover supersymmetric particles
and probe their properties, one could  predict the dark matter
density of the Universe and would  constrain   cosmology with the
help of precision data provided by WMAP and PLANCK.
  We investigate how well the relic density can be predicted in
  minimal supergravity (mSUGRA), with and without the assumption of mSUGRA
  when analysing data.
  We determine the parameters to which the relic density is most sensitive,
  and    quantify the collider accuracy needed. Theoretical errors in the
  prediction   are investigated in some detail.
}


\preprint{LAPTH-1070/04\\  DAMTP-2004-100\\ hep-ph/0410091}

\begin{document}

\newcommand{\beqn}{\begin{eqnarray}}
\newcommand{\eeqn}{\end{eqnarray}}
\newcommand{\ra}{\rightarrow}

%% \newcommand{\np}{Nucl.\,Phys.\,}
%% \newcommand{\pl}{Phys.\,Lett.\,}
%% \newcommand{\pr}{Phys.\,Rev.\,}
%% \newcommand{\prl}{Phys.\,Rev.\,Lett.\,}
%% \newcommand{\prep}{Phys.\,Rep.\,}
%% \newcommand{\nuclinst}{{\em Nucl.\ Instrum.\ Meth.\ }}
%% \newcommand{\annp}{{\em Ann.\ Phys.\ }}
%% \newcommand{\intjmp}{{\em Int.\ J.\ of Mod.\  Phys.\ }}

%GENERAL

\newcommand{\mw}{M_{W}}
\newcommand{\mww}{M_{W}^{2}}
\newcommand{\mwmw}{M_{W}^{2}}

\newcommand{\mz}{M_{Z}}
\newcommand{\mzz}{M_{Z}^{2}}

\newcommand{\cw}{\cos\theta_W}
\newcommand{\sw}{\sin\theta_W}
\newcommand{\tw}{\tan\theta_W}
\def\gev{\mbox{GeV}}
\def\cww{\cos^2\theta_W}
\def\sww{\sin^2\theta_W}
\def\tww{\tan^2\theta_W}

\def\noi{\noindent}
\def\nn{\noindent}

\def\sinb{\sin\beta}
\def\cosb{\cos\beta}
\def\sinbb{\sin (2\beta)}
\def\cosbb{\cos (2 \beta)}
\def\tgb{\tan \beta}
\def\tgbt{$\tan \beta\;\;$}
\def\tgbsq{\tan^2 \beta}
\def\sel{\tilde{e}_L}
\def\ser{\tilde{e}_R}
\def\msel{m_{\sel}}
\def\mser{m_{\ser}}
\def\mslr{m_{\tilde{l}_R}}
\def\m0{m_0}
\def\amu{\delta a_\mu}

\def\slashE{E\kern -.620em {/}}
%%%%%%%%%%%%%%%%%%%%%

\def\mchi{m_\chi^+}
\def\neuto{\tilde{\chi}_1^0}
\def\mneuto{m_{\tilde{\chi}_1^0}}
\def\neutt{\tilde{\chi}_2^0}
\def\neutth{\tilde{\chi}_3^0}
\def\mneutth{m_{\tilde{\chi}_3^0}}
\def\ma{M_A}
\def\mstau{m_{\tilde\tau}}
\def\msne{m_{\tilde\nu}}
\def\msnee{m_{{\tilde\nu}_e}}
\def\mh{M_h}

\def\sinb{\sin\beta}
\def\cosb{\cos\beta}
\def\sinbb{\sin (2\beta)}
\def\cosbb{\cos (2 \beta)}
\def\tgb{\tan \beta}
\def\tgbt{$\tan \beta\;\;$}
\def\tgbsq{\tan^2 \beta}
\def\sinal{\sin\alpha}
\def\cosal{\cos\alpha}
%%%%%%%%%%%%%%%%%%%%%%%%5
\def\stop{\tilde{t}}
\def\sto{\tilde{t}_1}
\def\stt{\tilde{t}_2}
\def\stl{\tilde{t}_L}
\def\str{\tilde{t}_R}
\def\msto{m_{\sto}}
\def\mstosq{m_{\sto}^2}
\def\mstt{m_{\stt}}
\def\msttsq{m_{\stt}^2}
\def\mt{m_t}
\def\mtsq{m_t^2}
\def\sintsq{\sin^2\theta_{\stop}}
\def\costsq{\cos^2\theta_{\stop}}
\def\mqtt{\M_{\tilde{Q}_3}^2}
\def\mutt{\M_{\tilde{U}_{3R}}^2}
%%%%%%%%%%%%%%%%%%%%%
\def\sbottom{\tilde{b}}
\def\sbo{\tilde{b}_1}
\def\sbt{\tilde{b}_2}
\def\sbl{\tilde{b}_L}
\def\sbr{\tilde{b}_R}
\def\msbo{m_{\sbo}}
\def\msbosq{m_{\sbo}^2}
\def\msbt{m_{\sbt}}
\def\msbtsq{m_{\sbt}^2}
\def\mt{m_t}
\def\mtsq{m_t^2}
%%%%%%%%%%%%%%%%%%%%%
\def\selectron{\tilde{e}}
\def\seo{\tilde{e}_1}
\def\set{\tilde{e}_2}
\def\sel{\tilde{e}_L}
\def\se1{\tilde{e}_1}
\def\ser{\tilde{e}_R}
\def\mseo{m_{\seo}}
\def\mseosq{m_{\seo}^2}
\def\mset{m_{\set}}
\def\msetsq{m_{\set}^2}
\def\msel{m_{\sel}}
\def\mser{m_{\ser}}
\def\mse1{m_{\se1}}
\def\me{m_e}
\def\mesq{m_e^2}
%%%%%%%%%%%%%%%%%%%%%
\def\snu{\tilde{\nu}}
\def\snue{\tilde{\nu_e}}
\def\set{\tilde{e}_2}
\def\snul{\tilde{\nu}_L}
\def\msnue{m_{\snue}}
\def\msnuesq{m_{\snue}^2}
%%%%%%%%%%%%%%%%%%%%%
\def\smuon{\tilde{\mu}}
\def\smul{\tilde{\mu}_L}
\def\smur{\tilde{\mu}_R}
\def\msmul{m_{\smul}}
\def\msmulsq{m_{\smul}^2}
\def\msmur{m_{\smur}}
\def\msmursq{m_{\smur}^2}
%%%%%%%%%%%%%%%%%%%%%%%%%%
\def\stau{\tilde{\tau}}
\def\stauo{\tilde{\tau}_1}
\def\staut{\tilde{\tau}_2}
\def\staul{\tilde{\tau}_L}
\def\staur{\tilde{\tau}_R}
\def\mstauo{m_{\stauo}}
\def\mstauosq{m_{\stauo}^2}
\def\mstaut{m_{\staut}}
\def\mstautsq{m_{\staut}^2}
\def\mtau{m_\tau}
\def\mtausq{m_\tau^2}
\def\sint{\sin\theta_{\stau}}
\def\sintt{\sin 2\theta_{\stop}}
\def\cost{\cos\theta_{\stau}}
%%%%%%%%%%%%%%%%%%%%%%%%%%%%%%
\def\gluino{\tilde{g}}
\def\mgluino{m_{\tilde{g}}}
\def\mchi{m_\chi^+}
\def\neuto{\tilde{\chi}_1^0}
\def\mneuto{m_{\tilde{\chi}_1^0}}
\def\neutt{\tilde{\chi}_2^0}
\def\mneutt{m_{\tilde{\chi}_2^0}}
\def\neutth{\tilde{\chi}_3^0}
\def\mneutth{m_{\tilde{\chi}_3^0}}
\def\neutf{\tilde{\chi}_4^0}
\def\mneutf{m_{\tilde{\chi}_4^0}}
\def\chargop{\tilde{\chi}_1^+}
\def\chargopm{\tilde{\chi}_1^\pm}
\def\mchargo{m_{\tilde{\chi}_1^+}}
\def\chargtp{\tilde{\chi}_2^+}
\def\mchargt{m_{\tilde{\chi}_2^+}}
\def\chargom{\tilde{\chi}_1^-}
\def\chargtm{\tilde{\chi}_2^-}
\def\bino{\tilde{b}}
\def\wino{\tilde{w}}
\def\photino{\tilde{\gamma}}
\def\zino{tilde{z}}
%%%%%%%%%%%%%%%%%%%%%%%%%%%%%%%%%
\def\sdowno{\tilde{d}_1}
\def\sdownt{\tilde{d}_2}
\def\sdownl{\tilde{d}_L}
\def\sdownr{\tilde{d}_R}
\def\supo{\tilde{u}_1}
\def\supt{\tilde{u}_2}
\def\supl{\tilde{u}_L}
\def\supr{\tilde{u}_R}
%%%%%%%%%%Higgses masses%%%%%%%%%%%%
\def\mb{m_b}
\def\mhf{M_{1/2}}
\def\dMb{\Delta m_b}
\def\mbmb{m_b(m_b)}
\def\tb{\tan \beta}
\def\mh{m_h}
\def\mht{m_h^2}
\def\MH{M_H}
\def\MHt{M_H^2}
\def\MA{M_A}
\def\MAt{M_A^2}
\def\MHp{M_H^+}
\def\MHm{M_H^-}
\def\epem{e^+e^-}
\def\epemt{$e^+e^-$}
\def\siginv{\sigma_{\gamma+inv}}
\def\gmuon{$(g-2)_\mu$}
\def\r12{r_{12}}
\def\bsgamma{b\ra s \gamma}
\def\bsmu{B_s\ra \mu^+\mu^-}
\def\feynhiggs{{\tt FeynHiggs}}
\def\micro{{\tt micrOMEGAs}}
\def\mhf{M_{1/2}}
\def\suspect{{\tt Suspect}}
\def\softsusy{{\tt SOFTSUSY}}
\def\xenon{{\tt Xenon}}
\def\edelweiss{{\tt Edelweiss}}
\def\genius{{\tt Genius}}
\def\cdms{{\tt CDMS}}
\def\picasso{{\tt PICASSO}}

\section{Introduction}

One of the attractive features of the minimal supersymmetric
standard model (MSSM) is that it provides a natural candidate for
cold dark matter, the neutralino $\neuto$~\cite{Ellis:1983ew}.
With cosmology entering the era of precision measurements and the
next colliders aiming at discovering and constraining
supersymmetry some crucial cross breeding is emerging. Already,
assuming the standard cosmology~\cite{Kolb:1990vq}, the
measurement of the relic density of dark matter has been used to
put strong constraints on the supersymmetric
model~\cite{Baer:2003yh,Ellis:2003cw,Chattopadhyay:2003xi,Lahanas:2003yz}.
For example, WMAP~\cite{Spergel:2003cb,Bennett:2003bz}, which at
$2\sigma$ constrains the relic density in the range $.094<\Omega
h^2<.129$, effectively reduces the dimensionality of the MSSM
parameter space by one. $\Omega$ is the mass density in units of
the critical density and $h$ is the scaled Hubble constant. The
accuracy of this constraint is expected to increase with future
data from the PLANCK satellite, which should obtain precision on
$\Omega h^2$ at the 2$\%$
level~\cite{Lesgourgues:1998sg,Balbi:2003en}.

Previously, much of the literature has used $\Omega h^2$ as well
as other measurements in order to place constraints upon the
MSSM~\cite{Ellis:2003cw,Chattopadhyay:2003xi,Lahanas:2003yz,Baer:2002gm,Belanger:2004ag,Baltz:2004aw,Battaglia:2003ab},
sometimes giving predictions for the signatures expected at
colliders and direct dark matter searches. In this paper, we look
at the problem from the inverse perspective and examine what is
required from collider data in order to get a precise prediction
for $\Omega h^2$, which could then be used to test the cosmology.
For example, we can ask: is {\em all} of the dark matter due to
the lightest neutralino, or is there some other component?
Assuming that the neutralino constitutes all of the dark matter,
the predicted $\Omega h^2$ value for neutralinos may test the
cosmological assumptions that go into the derivation of the relic
density in the standard picture. A value predicted from colliders
with enough accuracy that does not match with the WMAP number
would indicate some non-standard cosmology. This would be
highly exciting.\\
In the standard picture, at early times, the would be dark matter
relics are in thermal equilibrium with a number density $n_{eq}\propto
(M T)^{3/2} exp(-M/T)$ for a non-relativistic particle of mass $M$
at temperature $T$. Once the interaction rate of the particles
falls below the rate of expansion of the Universe set by the
Hubble parameter, the particles freeze-out with a number density
determined by the so-called freeze-out temperature $T_f$, with
$T_f \sim M_f/25$. Considering the exponential decrease of
$n_{eq}$ the relic density therefore depends critically on the
precise moment the annihilation rate equals the expansion rate of
the universe. This interaction rate and its  temperature/time
evolution is completely controlled by particle physics. On the
other hand, in the standard approach one implicitly assumes that
the post-inflation era is radiation dominated. There is no
clear-cut evidence that between the freeze-out (at $T \sim
10-100$GeV, say) and big bang nucleosynthesis BBN (at $T \sim
1MeV$) the universe was radiation dominated. Radiation domination
could start at a ``reheat" temperature that may be quite low. If
the expansion of the universe is modelled differently than in the
standard picture so that some  other contributions or  dynamics
occur,  the link between particle physics and cosmology is no
longer unique. As examples, one can mention a few explicit
scenarios such as a low reheating temperature
~\cite{Giudice:2000ex,Khalil:2002mu}, scalar field
kination\cite{Salati:2002md,Rosati:2003yw,Joyce:1997cp,Joyce:2000ag},
scenarios with extra dimensions\cite{Nihei:2004xv}, anisotropic
cosmology ~\cite{Kamionkowski:1990ni,Profumo:2004ex}, or
scalar-tensor cosmologies~\cite{Catena:2004ba}. Entropy creation
as well as non-thermal production of
neutralinos~\cite{Murakami:2000me} can also change the picture.

For the particle physics part, in order to predict the relic
density $\Omega h^2$ in the MSSM one generally needs to know many
of the numerous  underlying parameters of the model, since that
determines the variety of  annihilation channels for neutralinos
in the early universe. Indeed, in the standard approach $\Omega
h^2 \propto 1/<\sigma v>$, where $<\sigma v>$ is the thermally
averaged cross section times the relative velocity of the LSP. How
many parameters could be known and with which precision depend
both on the theoretical model and the future terrestrial data,
especially from the colliders. At future colliders such as the
Large Hadron Collider (LHC) and a future linear collider facility
(LC), supersymmetric sparticles are expected to be produced and
measured if low energy supersymmetry (SUSY) is present in
nature~\cite{Branson:2001ak,Aguilar-Saavedra:2001rg}. It is hoped
(though not guaranteed) that enough
information~\cite{Allanach:2001qe,Allanach:2004my} will be present
in order to discriminate between various models of supersymmetry
breaking. When we have identified a successful model of SUSY
breaking, an accurate prediction of $\Omega h^2$ will allow us to
test the cosmological assumptions that go into its prediction and
to determine whether {\em all} of the dark matter is due to the
neutralino. This identification of the underlying SUSY breaking
scenario, although highly desirable, may not always be possible.
However to predict the relic density it may also happen that one
does not need to know all of the parameters but only those {\em
relevant parameters} that are needed for the relic density
calculation. An example of what we mean by an underlying SUSY
breaking scenario is the  minimal supergravity,
mSUGRA\footnote{Often referred to elsewhere as the CMSSM.} model.
Even when an underlying model has been identified, at the LHC for
example, the properties of those particles which play a dominant
r\^ole in the relic density calculation are not measured directly or
precisely.  The {\em relevant parameters} are then
inferred through a complex theoretical framework that brings with it an
associated uncertainty.

One aim of this paper is to critically examine the theoretical
uncertainties~\cite{Allanach:2003jw,Allanach:2004rh} in the
calculation of the MSSM spectrum and their impact on a precise
determination of the relic density as well as deriving the
accuracy needed on the fundamental parameters of the underlying
model. We also take the model independent approach where we
aim at only identifying the set of {\em relevant parameters}
which are the most important parameters to measure in different
regions of parameter space, and what empirical accuracy upon them
will be required. We will refer throughout to the WMAP benchmark
on accuracy: those that produce a 10$\%$ change in $\Omega h^2$
(``WMAP accuracy''). In part we also discuss what is needed to
achieve  the 2$\%$ level (``PLANCK accuracy''). Here, we will identify
requirements on observables more generally (along lines in
WMAP-allowed parameter space) and although we leave speculations
about the feasibility of meeting them to the conclusions, we
discuss which additional collider observables are worth
investigating to improve on the precision of the {\em relevant
parameters}. In this article we carry this investigation in the
mSUGRA framework with the supersymmetric spectrum provided by {\tt
SOFTSUSY1.8.7}~\cite{Allanach:2001kg}. The relic density is
computed with {\tt micrOMEGAs1.3}~\cite{Belanger:2004yn} which is
interfaced with {\tt SOFTSUSY1.8.7}  through the {\em SUSY Les
Houches Accord}~\cite{Skands:2003cj}.

The emphasis upon corroborating accurate cosmological measurements
with terrestrial collider data, rather than cosmology constraining
SUSY, has been addressed very recently, see for
example\cite{Boudjema:2001ni}. Ref.~\cite{Brhlik:2000dm}
investigated how the variation of the input parameters affected
the prediction of the relic density (and direct detection) but
this was a pre-WMAP study in a scenario no longer viable and, more
importantly, does not address the sensitive issues of poles and
co-annihilations that we will study. Ref.~\cite{Drees:2000he}
examines how, at the LHC, one could determine some parameters
entering the relic density calculation, while
Ref.~\cite{Polesello:2004qy} determines the required accuracy on
the fundamental parameters of mSUGRA before translating this into
an error on the relic density. Again both these studies were
pre-WMAP studies and hence in a region of parameter space which is
quite favourable for the LHC.   Some recent experimental
simulations covering part of the mSUGRA parameter space compatible
with  WMAP have been performed in
\cite{Bambade:2004tq,Martyn:2004jc} for the LC and in
\cite{Battaglia:2003ab} for the LHC.

The paper is organised as follows. We will start in
Section~\ref{sec:msugra} by briefly describing the three regions
in the mSUGRA parameter space that are still compatible with the
WMAP constraint and which we will study in detail in the next
sections. These are the $\stau$ co-annihilation region, the Higgs
funnel region and the focus point region. We felt that a dedicated
Section~\ref{sec:set-up},  devoted to how we quantify the
sensitivities and how we approach the problem, both purely within
mSUGRA and in a more model independent way, was warranted.  We
investigate the co-annihilation region first in
Section~\ref{sec:coan}. Results for the Higgs funnel region are
shown in Section~\ref{sec:largetb}. In section~\ref{sec:focus},
the focus point regime is studied.
 Conclusions follow in
section~\ref{sec:conc}. Input parameters for the analysis are
detailed in Appendix~\ref{sec:sminp}.

\section{mSUGRA post WMAP}
\label{sec:msugra}

mSUGRA is defined through only four parameters and a sign:
$M_{1/2},m_0,A_0$ i.e.\ the common gaugino mass, scalar mass and
tri-linear coupling respectively (all defined at the GUT scale) as
well as $\tgb$ the ratio of the vacuum expectation values in each
of the Higgs doublet at the weak scale. The sign refers to the
sign of the derived $\mu$ parameter, the Higgsino mass term. Just
prior to WMAP, mSUGRA was compatible with the limit $\Omega
h^2\sim 0.1-0.3$ and other direct and indirect low-energy and
collider data in a rather large region of parameter space called
the ``bulk region''. The SPS1a~\cite{Allanach:2002nj} mSUGRA
benchmark point lies in this region. It  was studied in an ATLAS
simulation with the assumption that nature follows
mSUGRA\cite{Polesello:2004qy}. It was concluded that a precision
of 3$\%$ on the relic density was possible in this case. After the
WMAP constraint upon $\Omega h^2$ however, and given empirical
lower bounds on sparticle masses, this large bulk region in
$m_0-M_{1/2}$ is no longer viable, see for
example~\cite{Belanger:2004ag}.  SPS1a   is now ruled out by the
WMAP relic density constraint.  The point is that in mSUGRA, the
neutralino LSP  happens to be, for practically all cases, an
almost pure bino that annihilates most efficiently into leptons
through the right-handed sleptons because of their larger
hypercharge assignment. However with the newest WMAP data this
mechanism is not efficient enough. There remains then three
favoured scenarios which all require some very specific accidental
relations between some parameters at the electroweak scale.

\begin{itemize}
\item{}$\stau$ co-annihilation:\\
In mSUGRA at small $m_0$, there exists a region with almost
degenerate $\stau-\neuto$. In this case  the population of these
two sparticles is almost the same, making the NLSP $\stau$
thermally accessible. $\Delta M=m_{\stau} - m_{\neuto}$ is the mass
difference which controls the ratio of the population of the two
species through the Boltzmann factor $\exp(-\Delta
M/T_f)$\cite{Griest:1991kh}. It is then a very sensitive parameter
that enters the calculation of the relic density. When
co-annihilation takes place, through the participation of the
$\stauo$ in processes such as $\stau_1
  \neuto
  \rightarrow \tau\gamma$ or even  $\stau_1
  \stau_1
  \rightarrow \tau \bar \tau$, the relic density can be brought
  down compared to the case of the bulk scenario.

\item{} Higgs funnel region: \\
A sudden increase in the usual annihilation mechanism to bring
down the relic density can also occur if $m_{\neuto}$ is near a
pole. Collider constraints on the LSP in mSUGRA  allow the heavy
``Higgs funnel''~\cite{Drees:1993am,Arnowitt:1993mg}: where
  $\neuto \neuto
  \rightarrow A   \rightarrow b \bar{b}/\tau \bar{\tau}$,
which occurs at large $\tan \beta$.
\item{} Focus region: \\
 Most of the time in mSUGRA, $\mu$ is rather large.
However it may exceptionally happen that $\mu \sim M_1$ or even
$\mu \lesssim M_1$, in which case the annihilation is much more
efficient through reactions such as $\neuto \neuto \rightarrow
WW/ZZ/Zh/t {\bar t}$. This occurs in the so-called focus point
region~\cite{Feng:1999zg,Feng:2000gh,Birkedal-Hansen:2002sx,Baer:2002fv}
where $m_0$ is very large.
\end{itemize}

Our study will cover these three rather constrained regions in the
mSUGRA parameter space in detail. For all the MSSM models
considered here, the LSP is a neutralino. It is therefore useful
to list its parameters through the neutralino mass matrix which
will help understand some of our results.

The neutralino mass matrix, in  the bino, wino, ``up" and "down"
Higgsino basis respectively is defined as
\beqn
\label{neutralinomatrix} \left(
\begin{array}{cccc}
M_1 & 0 & -\mz \sw \cosb & \mz \sw \sinb \\ 0 & M_2 & \mz \cw
\cosb & -\mz \cw \sinb \\ -\mz \sw \cosb   & \mz \cw \cosb  & 0 &
-\mu \\ \mz \sw \sinb & -\mz \cw \sinb & -\mu & 0
\end{array}  \right),
\eeqn
\noi where $M_1$ and $M_2$ correspond to the weak scale bino and
wino masses. $M_Z$ is the $Z$ mass. This matrix is diagonalised
with a matrix that we label $N$. In mSUGRA, with $\theta_W$ the
weak mixing angle one has
\beqn
\label{m1m1unification} M_1=\frac{5}{3} \tan^2\theta_W M_2 \simeq
M_2/2.
\eeqn

This matrix receives radiative corrections but keeps the {\em
same} form as the above tree-level structure. In our analysis
these radiative corrections consist of an approximation to the
full one-loop result, based on Ref.~\cite{Pierce:1996zz}. All
sparticle mixing is ignored in the correction terms, corrections
proportional to the $U(1)$ coupling are ignored, quark masses are
set to zero and the squarks are approximated to be degenerate.
Also, tree-level formulae for the Higgs are used in the loop
corrections to the mass matrix. The tau Yukawa coupling is
neglected. These approximations induce errors of order $(\alpha/4
\pi) M_Z^2 /
\mu^2$ and $(\alpha/4 \pi) M_Z^2 / M_{A}^2$. $M_A$ is the mass of
the pseudo-scalar Higgs.

\section{The set-up}
\label{sec:set-up}

\subsection{The approach}
 To ``generate" the models compatible with the WMAP data we
will fix  $A_0$, $\tgb$ and the sign of $\mu$ but scan over the
GUT value of the gaugino mass $M_{1/2}$. The strong constraint
from WMAP then means that $m_0$ can not be varied at will but, is
to a large extent, determined by $M_{1/2}$. We therefore first
seek a parameterisation of the $m_0$-$M_{1/2}$ dependence which
describes the three regions of the $m_0$-$M_{1/2}$ plane which we
are interested in. Each region will then  be defined by a slope.
Of course, in order to arrive at the full spectrum at the
electroweak scale that is needed for the calculation of the
various  rates entering the relic density, one also needs to
specify the standard model parameters, such as $m_b(m_b), m_t$
etc.To derive the weak scale parameters we rely  on {\tt
SOFTSUSY}. {\tt SOFTSUSY} includes the necessary RGEs
(Renormalisation Group Equations) and calculation of pole masses.
It is well known that this procedure necessarily introduces some
theoretical errors on the prediction of the weak scale physical
parameters depending on the level of sophistication (loop order)
and approximations that are implemented in the code. In a
preliminary analysis we have compared different codes for the
spectrum evaluation and their impact on the prediction of the
relic density\cite{Allanach:2004jh}. Here, for all three cases, we
estimate the theoretical uncertainty solely with {\tt SOFTSUSY}.
For example, {\tt SOFTSUSY} uses a renormalisation scale
$M_{SUSY}$ at which the electroweak symmetry breaking conditions
are imposed, and at which most of the sparticle masses are
calculated. By default, $M_{SUSY}=\sqrt{m_{{\tilde t}_1}
m_{{\tilde t}_2}}$ is the renormalisation scale used at which it
is hoped sensitivity to higher order corrections is small. In
principle, changing this scale only has an effect on physical
observables at a perturbative order higher than that used in {\tt
  SOFTSUSY}. The {\em present}
theoretical uncertainty coming from higher order effects is then
estimated by re-calculating $\Omega h^2$ for different values of
\begin{equation}
x \equiv M_{SUSY}' /  M_{SUSY}. \label{x}
\end{equation}
Here, we take $x=0.5-2.0$ to represent the scale variation. We
will, depending on the situation, also study other uncertainties
stemming from the different values of the SM parameters.
Some of the theoretical uncertainty in {\tt \small SOFTSUSY} could be 
reduced by using the full two-loop effective electroweak potential. Such a 
potential has been presented in Ref.~\cite{martin1} and was shown to 
reduce the scale dependence of the one-loop calculation for $\mu$ and $B$.
In Ref.~\cite{martin2}, the 2-loop strong and Yukawa corrections were 
shown to have a similar stabilising effect upon the Higgs pole masses. 
The spectrum generator {\tt \small SOFTSUSY1.8.7} used here includes the 
dominant terms of this two-loop subset in both the effective electroweak 
potential and the calculation of the neutral Higgs boson masses. Terms of 
order $h_t^4$, $h_t^2 g_3^2$, $h_b^2 g_3^2$, $h_b^4$, $h_b^2
h_t^2$, $h_\tau^4$, $h_\tau^2 h_b^2$ are all included. 
It will clearly be desirable to include other sub-dominant terms from 
Refs.~\cite{martin1,martin2} when they become available in a form easy to 
include in the spectrum generator.

This preliminary investigation  sets the stage as concerns the
improvements, both theoretical and experimental, that are needed
if one aims at a prediction of the relic density starting from the
mSUGRA (high scale) parameters. Otherwise, some of these
uncertainties should be taken into account
as additional systematic errors. \\
\noi We then address the question of how precisely, purely within
mSUGRA, do we need to reconstruct the defining model parameters
$m_0,M_{1/2},\tgb,A_0$ to meet the accuracy of WMAP, say. We will
give below the detailed procedure of how we arrive at the required
accuracy on these parameters. This  model dependent approach is of
relevance for analyses at the colliders, especially LHC analyses,
where the parameters can be extracted from a global fit to a
number of observables\cite{Armstrong:1994it,Allanach:2003jw}. In
some instances, the standard model input parameters  such as the
top mass or the strong coupling are crucial because of RGE
effects. We also investigate the required accuracy on these
parameters. Let us mention in this respect that the uncertainties
in $m_b(m_b)$ were investigated in ref.~\cite{Gomez:2004ek} for
the funnel region.\\
\noi In a second approach, which for short we will call PmSUGRA,
we aim at a less model dependent approach. The spectrum is still
generated within mSUGRA in one of the three regions in accord with
WMAP. Here however we aim at finding out the accuracies on the
physical parameters at the electroweak scale. This may seem a
daunting task since the calculation of the relic density involves
quite a few channels\footnote{To cover the most general case, {\tt
micrOMEGAs}\cite{Belanger:2001fz,Belanger:2004yn} includes some
$3000$ channels or so.} which in turn means having access to the
entire parameter space of the general MSSM. The fact is that, for
the three regions under study, the relic density is controlled by
a quite small set of physical parameters and is rather insensitive
to the rest of the physical parameters. We foresee this to be the
case in most SUSY scenarios constrained by WMAP. We therefore
start by concentrating on these relevant weak scale parameters. To
help pin down these parameters we will show, along the slope, the
relative contribution of those most important cross sections that
are necessary to predict the relic density. It is educative at
this point to characterise how the relative importance of the
different channels comes about. For the co-annihilation channels,
the Boltzmann factor is, most of the time, an overwhelming factor.
This thermodynamical term requires a rather precise knowledge of
the mass difference between the LSP and the NLSP. Analyses at
colliders, in particular the LHC, may have direct access to this
mass difference rather than on the individual masses.
Knowledge of the cross sections  requires the knowledge of
couplings. Some of these are set not only by SM couplings but they
also involve for example mixing in the neutralino and stau sector.
In the funnel regions, mass measurements are again crucial due to
the resonant structure of these contributions. Anyway, in this
PmSUGRA approach we will, after identifying the relevant
parameters, seek to find out the accuracy needed. \\

We will generally refrain from addressing the issue of how either
the GUT scale parameters or the relevant physical parameters in
the PmSUGRA approach can be extracted in a global fit, or
otherwise, from physical observables at the colliders. Our main
concern is to derive the required accuracy needed to match WMAP
(or PLANCK).

\subsection{Deriving the accuracies through an iterative
procedure}

We ask the question: what fractional change $a=|\Delta p/p|$ in an
input parameter $p$ results in a fractional change $r=\Delta
\omega / \omega=10\%$ for WMAP accuracy? $\omega=\Omega_{\chi^0}
h^2$, for short. The PLANCK accuracy is obtained by dividing that
of WMAP by a factor of 5, when the WMAP accuracy is not too large
(below $50\%$, say), so we will usually
  decline to list it  for the purpose of brevity. We answer our question by
  an iterative
procedure\footnote{Taking the linear approximation and using $d \ln
\omega/ d \ln p$ results in large numerical errors in some cases.}. Let us
choose a point in parameter space and denote the resulting relic
density as $\omega_{-1}$. 
Taking an initial guess for $a_0$ of 0.1, for instance leads to a certain 
value $\omega_0$ for the relic density. Taking $i=0$ in the equation
\begin{equation}
a_{i+1} = a_i \frac{r}{(\omega_{i} - \omega_{-1}) /
\omega_{-1}}, \label{it}
\end{equation}
leads to a prediction for $a_{1}$. Here,
$\omega_i $ is the relic density that results from using the
parameter $p$ changed by a fractional amount $a_i$, so  we may
iterate Eq,~\ref{it} up to arbitrarily high $i$. 
For a given trial in the fractional change of the parameter $a_i$, the 
fractional factor in Eq.~\ref{it} is larger than one if $w_i$ comes out 
to be too small and smaller than one otherwise. 
As $i \rightarrow 
\infty$, $a_i \rightarrow a$. Thus, we apply Eq.~\ref{it} successively 
until $(a_{i+1} - a_i)/a_i < t$, where
$t$ is the numerical tolerance of the iteration (taken here to be
0.001). The resulting value of $a_i$ gives us a good approximation
to $|\Delta p/p|$.
We only examine the 
sensitivity to one parameter at a time and so the rest are all set at the 
values used for the relevant  point in parameter space being studied.
It should therefore be borne in mind that when quoted,
precisions on inputs are necessary but not always sufficient. For
example, the accuracy of several inputs must be constrained
simultaneously. There is a systematic in the calculation of
$\Omega h^2$, through {\tt micrOMEGAs}, that we have not taken
into account. The error in solving the Boltzmann equation could be
as high as $1\%$, not so important when compared to WMAP accuracy,
but vital when one goes to PLANCK accuracies. This error will
however be greatly improved by the time we have identified the
model as by then we will have a tailor made code, rather than a
general purpose code, for the calculation of the relic density.

We now turn to the definition of the sensitivity parameters in
both the mSUGRA and PmSUGRA approach we have just outlined.

\subsection{Analysis in the  mSUGRA approach}
We calculate $a(m_0), a(M_{1/2}), a(\tan \beta), a(\alpha_s(M_Z)),
a(m_b(m_b))$ by the iterative procedure described above: given an
initial mSUGRA point, we first change e.g. $m_0$, while
re-calculating the spectrum and couplings. We simultaneously
satisfy constraints coming from REWSB (Radiative Electroweak
Symmetry Breaking) conditions, experimental inputs on SM data and
the theoretical boundary condition on the high-scale SUSY breaking
parameters. We denote the new value $m_0'$, which is found by the
iterative procedure to induce a  fractional increase of 10$\%$ in
$\omega $. $a(m_0)=m_0'/m_0-1$ has analogous expressions for
$\mhf,\tan\beta$. When $a$ is calculated for other parameters,
$m_0$ is set back to its default value. $a(A_0)$ needs a slightly
different definition since we will be examining the case $A_0=0$.
To avoid division by zero in this case, we will simply calculate
which $\Delta A_0$ obtains the 10$\%$ fractional increase in
$\omega $, that is, the overall change rather than the fractional
one. For reasons detailed later in the text, we need to consider
$a(\Gamma_A)$, the total width of the pseudoscalar Higgs boson.
This is calculated by finding $a(m_b(m_b))$, which corresponds to
a certain $a(\Gamma_A)$. $\Gamma_A$ is sensitive to $m_b(m_b)$
through decays into bottom quarks since the Yukawa coupling is
proportional to it. So here the use of $\Gamma_A$ is traded off
for $m_b(m_b)$\footnote{This is done for technical reasons having
to do with the way $\Gamma_A$ is calculated and used in {\tt
micrOMEGAs}.}.

By default, we will consider 10$\%$ increases in $\omega$
($r=0.1$). There were cases when not all of the parameter space
considered allowed a 10$\%$ increase in $\omega $ by varying
certain parameters. For example, if the required inputs of
$\alpha_s(M_Z)$ or $m_t$ become too large, QCD becomes
non-perturbative or the top Yukawa coupling reaches a Landau pole
before the GUT scale. When this was the case, we chose to examine
the value of $a$ that results in a 10$\%$ decrease in $\omega$
($r=-0.1$). The sign of $r$ does affect the value of $a$ obtained
in some cases by a factor of up to 2. Whenever we have used
$r=-0.1$, $a(p)^-$ or $a(p^-)$ will denote the value of $a$ (ie a
superscript ``-''). When $a(p)>1$ for a particular parameter $p$,
we will often choose not to include that parameter in the plots
(unless it is expected to be  very difficult to measure) for
brevity.
%Let us note that the SUSY spectrum as given by {\tt
%SOFTSUSY} is always transmitted  to {\tt   micrOMEGAs} through the
%{\em SUSY Les Houches Accord} interface\cite{Skands:2003cj}.

\subsection{Analysis in the  PmSUGRA approach}
Here, we again pick a parameter point in mSUGRA derived from the
high-scale SUSY breaking terms  $m_0,\mhf,A_0$ and which is
compatible  with REWSB
and with the data on Standard Model particle masses and gauge
couplings. But when we calculate the required change $a(p)$ in the
parameter $p$ {\em we assume the more general MSSM}. As explained
above, this allows us to leave the strong mSUGRA assumption behind
when analysing data. The parameter $p$ can be a weak scale
parameter that defines the effective Lagrangian, examples being the
weak scale values of $M_{1,2},\mu, A_{\tau},
m_{\ser,\sel,\staur,..}$ beside the ubiquitous $\tgb$. These may
not be directly extracted from  physical observables at colliders,
however their advantage is that they set the physical masses {\em
and} the couplings of the SUSY particles through their mixing
matrices. We will, however, most often express the accuracies on
the {\em physical} parameters, such as  the physical masses. This
approach has a direct interpretation in terms of what is measured
at the colliders. On the other hand in some cases its
implementation for deriving the accuracies can prove tricky. For
example insisting that the neutralino mass matrix keeps its
tree-level form as is implemented in {\tt SOFTSUSY} means that one
can not vary one of the neutralino physical masses while freezing
the diagonalising matrices that define the couplings. Another
example concerns $\tgb$. For example, in the funnel region
most of the $\tgb$ dependence is
contained in the Yukawa coupling through
the pseudo-scalar Higgs exchange. Measuring $\Gamma_A$ will get
rid of the large $\tgb$ sensitivity. However this will not
translate into a measurement of $\tgb$ since $\Gamma_A$ depends on
other parameters. It is then interesting to ask how sensitive one
is to the $\tgb$ which enters the $A\neuto \neuto$ couplings, which
might be more difficult to extract experimentally.

%%%%%%%%%%%END OF SET UP %%%%%%%%%%%%%%%%%%%%%%%%%%%%

\section{Co-annihilation \label{sec:coan}}

\subsection{Characteristics of the co-annihilation sliver in
mSUGRA} In this region the LSP neutralino is mostly a bino with a
mass set essentially by $M_1$ up to corrections of order
$M_Z^2/\mu$ ($\mu$ is large). In the approximation where $\tgb$ is
large, in fact $\tgb>5$ will do, we may write :
\beqn
\label{eq:mneutco} \mneuto \simeq M_1 - \frac{M_Z^2}{4\mu^2} \left( M_1+\frac{2\mu}{\tgb}\right)
\quad {\rm and} \quad M_1=0.417M_{1/2}.
\eeqn
The mass of the right-sleptons in mSUGRA can  be approximated by
\beqn
\label{mer-co} m_{{\tilde l}_R}^2 \simeq m_0^2+0.152 M_{1/2}^2
\simeq m_0^2 +(0.390\mhf)^2.
\eeqn
In the same approximation, the mass of the left-sleptons may be
cast into
\beqn
m_{{\tilde l}_L}^2 \simeq m_0^2+0.52 M_{1/2}^2.
\eeqn
Therefore for a small enough $m_0$ one can easily understand how
$\mneuto \sim m_{{\tilde l}_R}$. Moreover, with high enough $\tgb$
and a larger $\tau$ Yukawa coupling, $m_{\stau_1}$ is even lighter
than the $\ser$ and the $\smur$. First $m_{{\tilde \tau}_R}<m_{{\tilde
    e}_R,{\tilde \mu}_R}$ through
RGE running effects involving third family Yukawa couplings. Second, $L-R$
mixing in the stau mixing matrix makes
$m_{\stau_1}< m_{{\tilde \tau}_R}$. $\stauo$ is the lightest of
the two staus and we define it in terms of the current eigenstates
with the mixing angle $\theta_{\stau}$ such that $\stauo=\cos
\theta_{\stau} \stau_R +
\sin
\theta_{\stau}
\stau_L$. Approximations for the $\stauo$ mass and the
mixing angle, neglecting the $A_\tau$ contribution, will also
prove useful. Using the weak scale parameters, in our scenario
we may write
\beqn
\label{mstauo-1} m_{\stauo}^2
\simeq m_{\tilde{\tau}_R}^2+m_\tau^2+\frac{M_Z^2}{4}
\left(1-\frac{2}{\tb^2}\right) -\frac{(m_\tau
\mu
\tgb)^2}{m_{\tilde{\tau}_L}^2-m_{\tilde{\tau}_R}^2}.
\eeqn
In mSUGRA this can be further simplified as
\beqn
\label{mstauo-2} m_{\stauo}^2
&\simeq& m_0^2 +0.152\mhf^2 -\epsilon_{{\rm RGE}}(\mu \tgb) +
\frac{M_Z^2}{4}
\left(1-\frac{2}{\tb^2}\right) - m_\tau^2 \left(\frac{
\mu
\tgb}{0.6\mhf}\right)^2,\nonumber \\
\tan 2 \theta_\tau&\simeq &\frac{2 m_\tau \mu \tgb}{(0.6
M_{1/2})^2} \simeq \frac{ m_\tau}{\mneuto}\frac{ \mu
\tgb}{\mneuto}.
\eeqn

As a measure of mixing we choose to display our results as a
function of $\cos 2 \theta_{\stau}$, since this quantity directly
enters the cross sections and decays involving the $\stau$'s.
$\epsilon_{RGE}(\mu \tgb)$ is the Yukawa contribution to the RGE
running.

\FIGURE{ \unitlength=1.3in
\begin{picture}(4,2.6)
\put(0.7,0){\epsfig{file=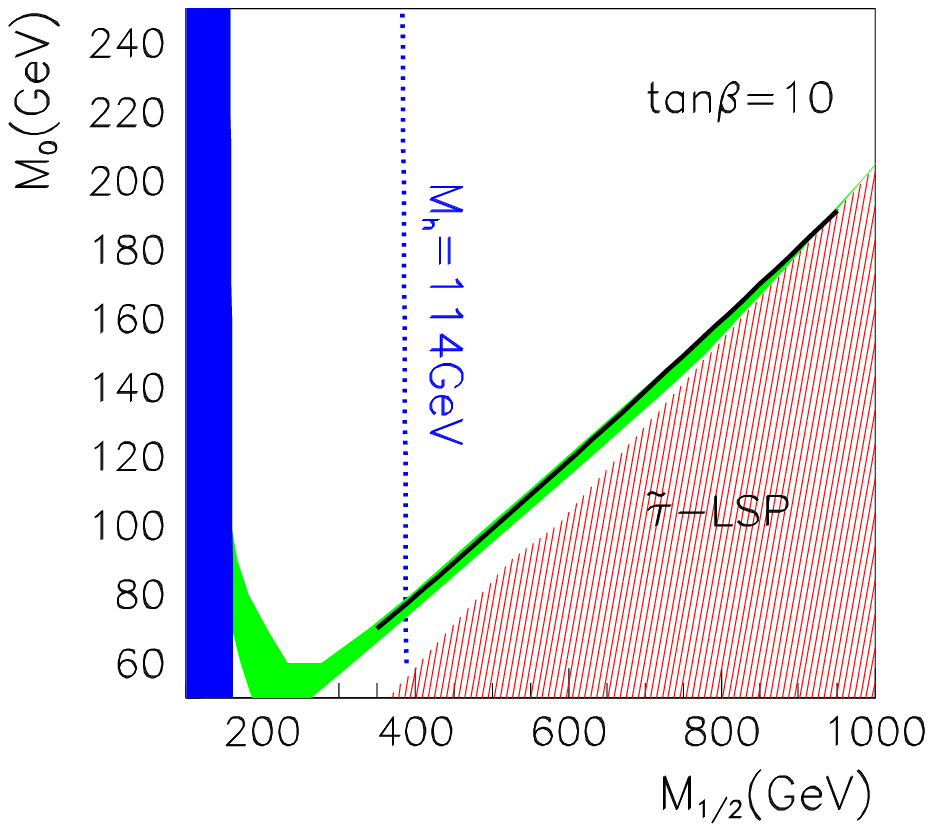, width=\wth}}
\end{picture}
\caption{Slope S1 in the $\m0-\mhf$ plane (full line) and the WMAP allowed region (green/light grey). The blue/dark grey area is excluded by direct LEP limit on sparticles, the dotted vertical line gives the Higgs mass limit from LEP2 (the region to the left of this line
is excluded). The red area  is 
ruled out by the cosmological constraint that the LSP is neutral.
}
  \label{fig:slope1}}

The approximate formulae for the neutralino mass and the $\stauo$
mass suggest that degeneracy occurs for $m_0\sim 0.145\mhf$. The
$m_0-\mhf$ constraint imposed by WMAP  can be made more precise.
We can define a slope that parameterises the $m_0-\mhf$ line where
co-annihilation occurs in accord with the WMAP data, see Fig.~\ref{fig:slope1}. We will specialise to the case $\mu>0$, $\tan
\beta=10$ and $A_0=0$. Following Ref.~\cite{Battaglia:2003ab}, we
take a slope ``S1'' in parameter space:
\begin{equation}
\label{slopes1} \frac{m_0}{\gev} =
5.84615 + 0.176374 
\frac{M_{1/2}}{\gev} +  1.97802\times 10^{-5}
\left(\frac{M_{1/2}}{\gev}\right)^2.
\end{equation}

\FIGURE{\twographs{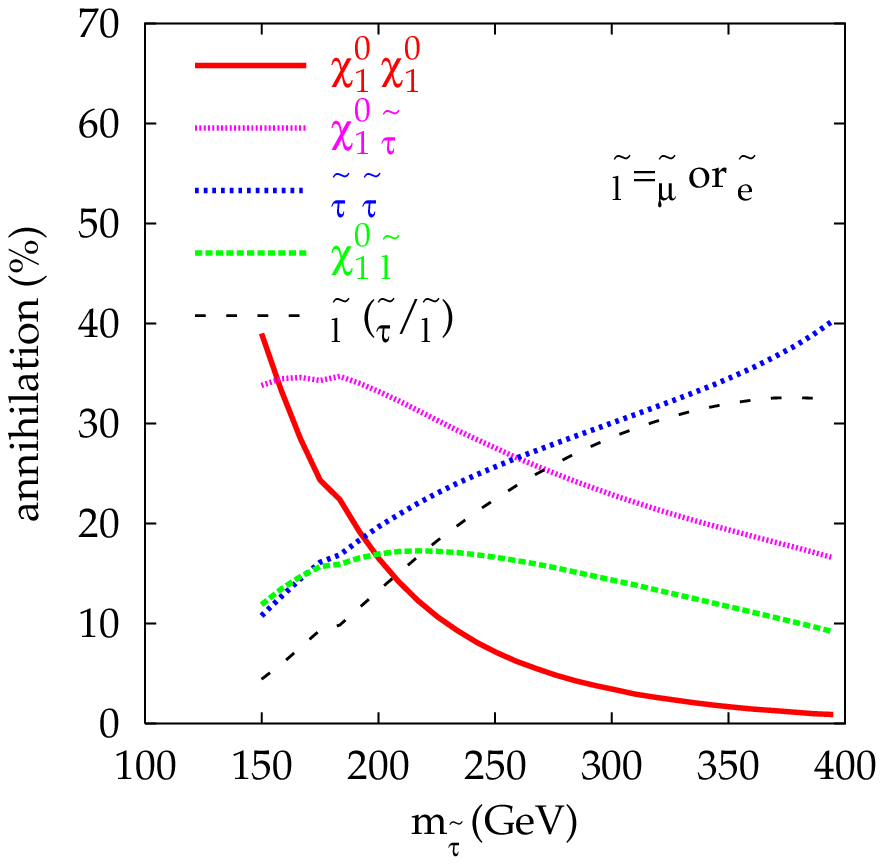}{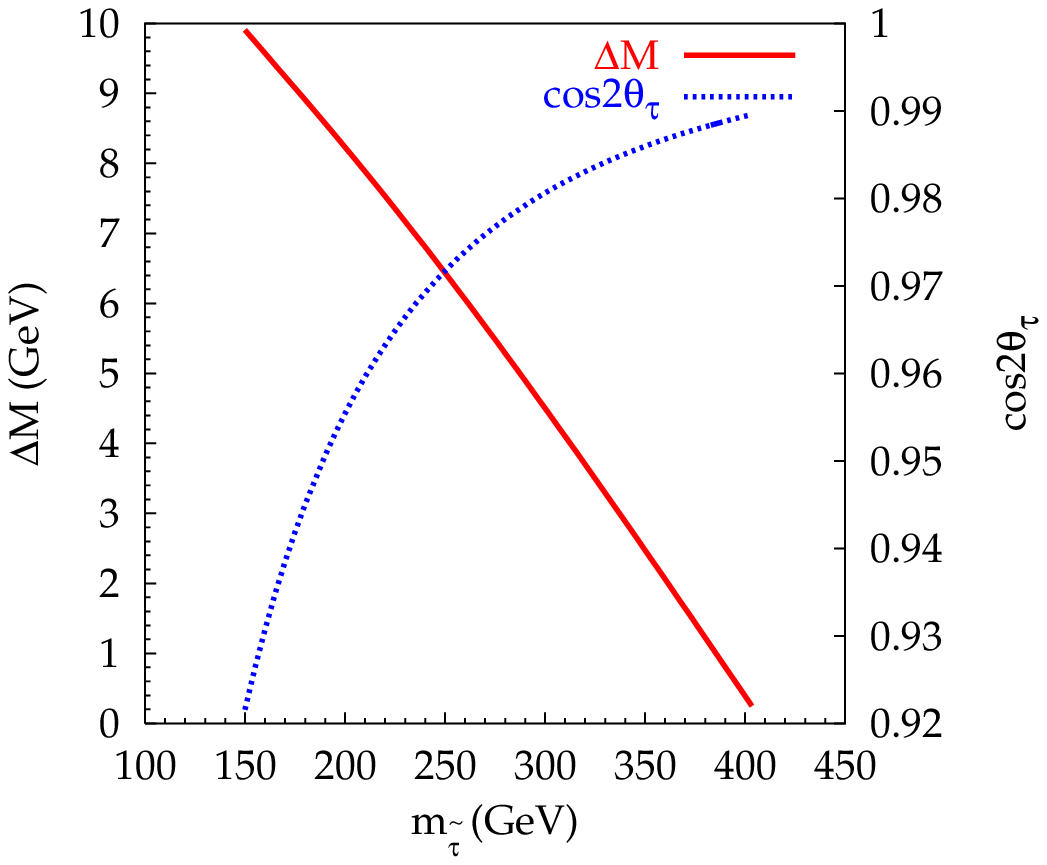} \caption{(a)
Contribution of the various channels to the relic density in $\%$.
(b) $\Delta M$ and $\cos\theta_{\stau}$ as a function of
$m_{\stau_1}$. The abscissa corresponds to $\mhf=350-920$ GeV.
(The label $\stau$ here stands in fact for $\stauo$.)
  \label{fig:coan-contrib}}}

This slope has a relic density in rough agreement with the WMAP
range. Along this slope, Fig.~\ref{fig:coan-contrib} shows the
relative contributions of the most important annihilation and
co-annihilations processes that allow to bring down the relic
density to the required WMAP level. The physical parameters that
enter the most important channels will be the most relevant
parameters that need to be measured precisely at the colliders.
The annihilation percentage in the figure corresponds to the proportion of 
neutralinos annihilated by a particular process. We
also note that the relative importance of some channels changes
considerably as the neutralino mass increases. It is important to
stress that although $\neuto \stauo$ co-annihilation is important,
it never makes up more than $50\%$ of the overall contribution.
For values of $\Delta M= \mneuto - m_{\stauo} \sim 10$GeV, it is
of the order of the annihilation contribution $\neuto
\neuto \ra {\rm all}$. The largest value of $\Delta M$ along the slope S1
displays the onset of the co-annihilation region, so one expects
the annihilation processes (usually associated to the bulk) not to
be negligible. As $\Delta M$ gets very small, of the order of a
few GeV, they are less important than $\stauo \stauo$ which then
dominates. With $\Delta M$ a few GeV, the co-annihilation channels
involving the smuon and selectron co-annihilations are also
important. Therefore independent measurement of the smuon and
selectron masses in model independent analyses will also be
required. It is important to note that as the mass scale $M$ (set
by the LSP mass, say) increases we require smaller and smaller
$\Delta M$. This can be understood from the fact that the cross
sections which scale as  $\propto 1/M^2$  are less and less
efficient and we require new channels to contribute. Past $M\sim
400$GeV this mechanism ceases to be
viable. \\
\FIGURE{ \unitlength=1.1in
\begin{picture}(4,2.6)
\put(0.7,0){\epsfig{file=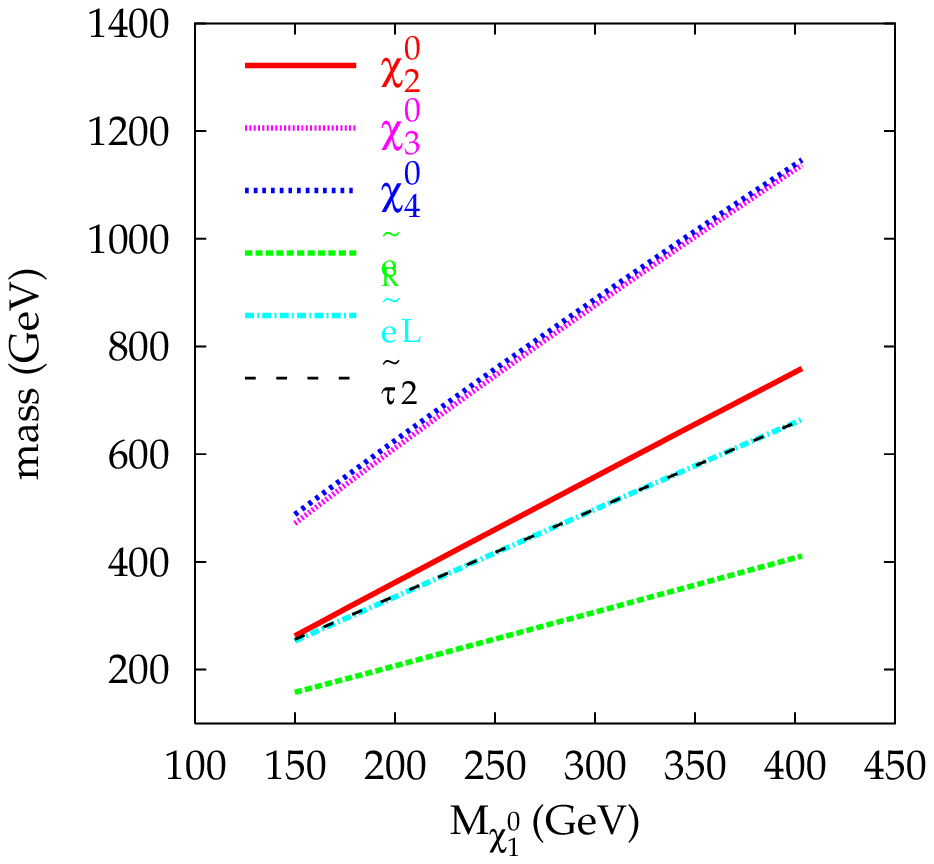, width=\wth}}
\end{picture}
\caption{Mass spectrum for some of the SUSY particles along the
$\stau$ co-annihilation slope.}
  \label{fig:coanSpec}}
\noi Fig.~\ref{fig:coan-contrib}(b) shows that $\Delta M$
decreases dramatically from $10$GeV to less than $1$GeV (that is,
less than the $\tau$ mass) for the higher $\stauo$ masses. Such
$\stau$'s will not have a two-body decay and might make the
reconstruction of some of the needed parameters a quite difficult
task. In particular, information extracted from the $\tau$
polarisation as suggested in \cite{Nojiri:1996fp} may be nearly
impossible to obtain, especially since we are never in a very
strong $\stau$ mixing scenario. Indeed, the mixing angle is small
throughout the slope but at high masses it is tiny. This behaviour
can be understood from Eq.~\ref{mstauo-2} for the mixing angle,
showing that indeed the mixing decreases with increasing LSP mass.
We also note that $\mu \sim m_{\chi^0_{3,4}} \sim 3 \mneuto$ on
the slope as shown in Fig.~\ref{fig:coanSpec}. This figure also
helps us to see which processes might be accessible at the
colliders. For instance at the onset of the co-annihilation
region, apart from the production of the three lightest sleptons
($\stauo, \smur, \ser$), there is a window for producing the
heavier sleptons as well as the associated $\neuto \neutt$ at a
$500$GeV \epemt machine, see also\cite{Arnowitt:2003za}. Of course
the maximum centre of mass energy will access more channels. It
may also be more judicious to perform a threshold scan for the
lightest stau as was done in \cite{Bambade:2004tq}, however since
we also want to access the smuon and selectron parameters a
combined analysis may prove more appropriate\cite{Martyn:2004jc}.
For the LHC it is worth mentioning that the cascade decay
$\tilde{q}\ra \neutt \ra
 \ser \ra \neuto$ will be available which can help in a good reconstruction
of the model parameters in the context of mSUGRA.

To close this subsection it is interesting to discuss the coupling
$\neuto \stauo \tau$. As seen from Fig.~\ref{fig:coan-contrib},
this coupling enters in a large number of processes. Defining its
right-handed part by $c_R$ and left-handed part by $c_L$ we have,
in the limit of moderate to large $\tgb$ we are considering:
\beqn
\label{neutstaucouplin}
c_R&=&-\frac{g}{\sqrt{2}}\left(2 \frac{s_W}{c_W} \cost N_{11}
+\sint
\frac{m_\tau \tgb}{M_W} N_{13} \right), \nonumber \\
c_L&=&\frac{g}{\sqrt{2}}\left( \sint (\frac{s_W}{c_W} N_{11}+
N_{12}) -
\cost
\frac{m_\tau \tgb}{M_W} N_{13} \right).
\eeqn
$N_{1j}$ are the elements of the matrix that diagonalises the
neutralino mass matrix, see Eq.~\ref{neutralinomatrix}. In the
scenario we are in with largish $\mu$, this becomes, at the
leading order,

\beqn
\label{neutstaucouplin-app}
c_R&=&-\frac{g}{\sqrt{2}} \frac{s_W}{c_W} \left(2  \cost +\sint
\frac{m_\tau \tgb}{\mu} \right), \nonumber \\
c_L&=&\frac{g}{\sqrt{2}} \frac{s_W}{c_W} \left(\sint -\cost
\frac{m_\tau \tgb}{\mu} \right).
\eeqn
Because of the small stau  mixing angle and because $\tgb$ is not
too large, the Yukawa contribution does not make too much of an
impact. We are therefore looking for a dependence on the mixing
angle coming from $\stauo \neuto \ra \tau \gamma$ essentially of
the form $3 \cos 2\theta_\tau +5$ from the cross sections.

\subsection{Theoretical uncertainties in mSUGRA}
\FIGURE{\twographs{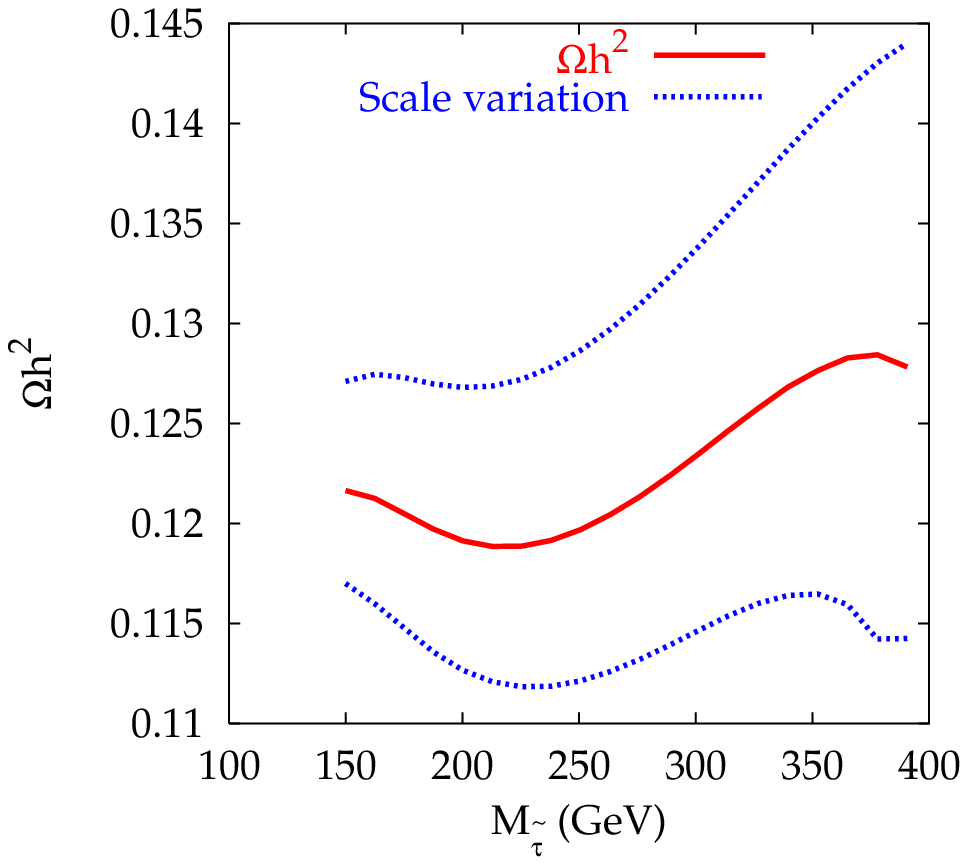}{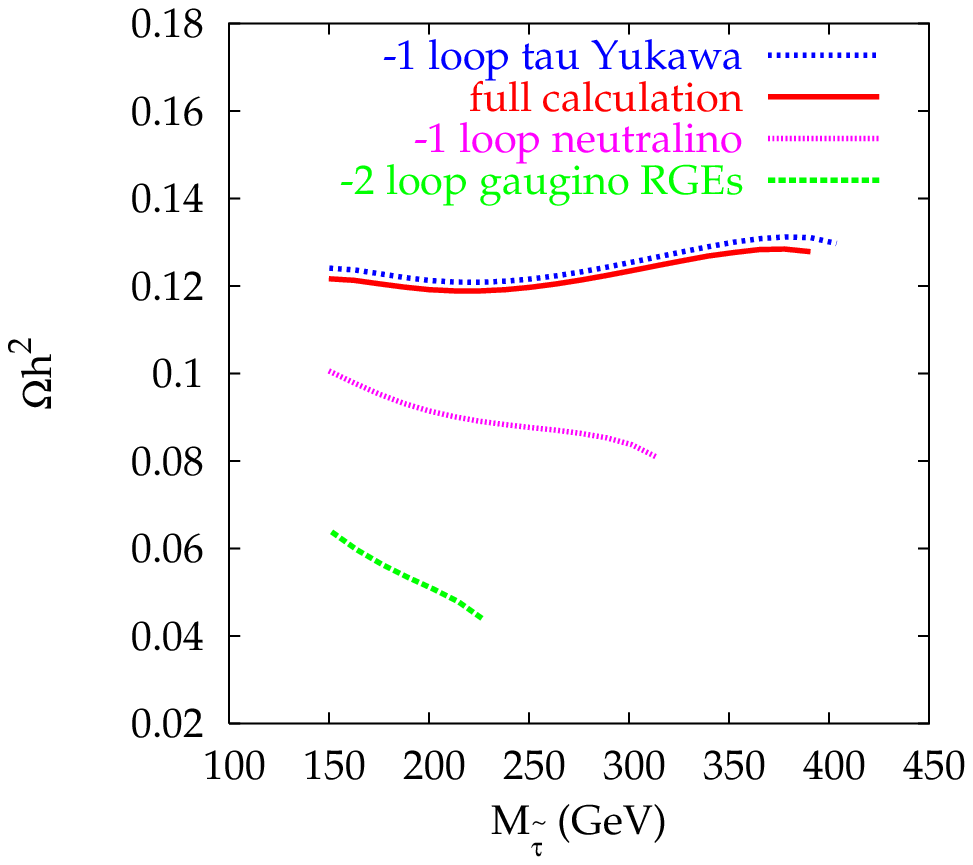}
\caption{(a) $\Omega h^2$ along slope S1. The central line
  gives the default {\tt SOFTSUSY} prediction, with the broken lines
  representing the limits due to the scale variation (as defined in the text).
(b) Effect of approximations along slope S1, the same default
$M_{{\rm SUSY}}$ scale is assumed here. We show the relic density
  for (top to bottom):
the default calculation
  neglecting SUSY corrections to the tau Yukawa coupling,
the default (``full'') calculation, the default calculation
neglecting
  loop threshold corrections to the neutralino mass,  and the default
  calculation  neglecting 2-loop terms in the renormalisation group
  equations  of gaugino masses.
The abscissa corresponds to $\mhf=350-920$ GeV.
  \label{fig:coan}}}

As pointed out earlier, predicting the relic density from the GUT
scale parameters is subject to theoretical uncertainties such as
the order at which the RGE are implemented. Fig.~\ref{fig:coan}a
(for $\mhf=350-920$ GeV, $m_0=73-185$ GeV) shows how the
prediction of the relic density changes as one varies the
renormalisation scale $M_{SUSY}$ at each point along the slope S1
by a factor 2 in each direction. In passing, observe that the
slope parameterises  the co-annihilation region rather well.
Controlling the scale uncertainty will be important in the purely
mSUGRA approach since these analyses may have to rely heavily on
the spectrum evaluator. The band of $\Omega h^2$ values found is
shown as the broken lines in Fig.~\ref{fig:coan}a. This theory
uncertainty is roughly $\pm 5\%$ at small $\mstau$, rising to $\pm
15\%$ at higher masses. This effect is due to the fact that
$M_{SUSY}$ varies over a much wider range at higher masses, also
that $\Delta M$ becomes very small and therefore fractionally more
sensitive to $M_{SUSY}'$.

In order to see what in the RGE or the calculation of the pole
masses from the $\overline{DR}$ is most important (and thus need to
be improved to deal more appropriately with the co-annihilation),
we look at the effect of some higher order loop corrections that
affect the neutralino and stau sector. We expect the next order of
presently non-calculated higher order corrections to have an
effect of two orders of magnitude below the removed corrections
(since they will be suppressed by a loop factor $\sim 1/100$). We
concentrate on approximations that have an appreciable effect upon
$\Omega h^2$.  In Fig.~\ref{fig:coan}b, we show the effect of the
important approximations along slope S1 by comparing the case when
all corrections are included, and when some are omitted. We see
from the figure that the two-loop terms in the RGEs of the gaugino
masses are essential to include, as are the one-loop threshold
corrections to the neutralino masses. 
Not including these contributions may even make $\mstauo>\mneuto$ and explain why some of the curves in Fig.~\ref{fig:coan}b do not extend as far as the curve corresponding to the full calculation.
Their effect is far greater
than the one induced by the SUSY scale. This could hint at the
necessity of including the three-loop gauge running and two-loop
threshold effects. Some work in this direction has been
performed\cite{Jack:2004ch} and the results seem to indicate that
at least for the LSP mass the three-loop running in the RGE has a
negligible effect. 
The threshold loop corrections to the tau Yukawa coupling
have a small effect. We were unable to study the effect of
threshold loop corrections to the stau masses, since {\tt
SOFTSUSY} does not yet include them\footnote{We plan to include
loop
  corrections to stau masses in a   future version of {\tt SOFTSUSY}.}.
According to ref.~\cite{Pierce:1996zz}, corrections to $m_{\tilde \tau}$ are
typically less than 1$\%$.

\subsection{Accuracies of the GUT scale mSUGRA parameters}

\FIGURE{
 \unitlength=1.1in
\begin{picture}(4,2.6)
\put(0.8,0){\epsfig{file=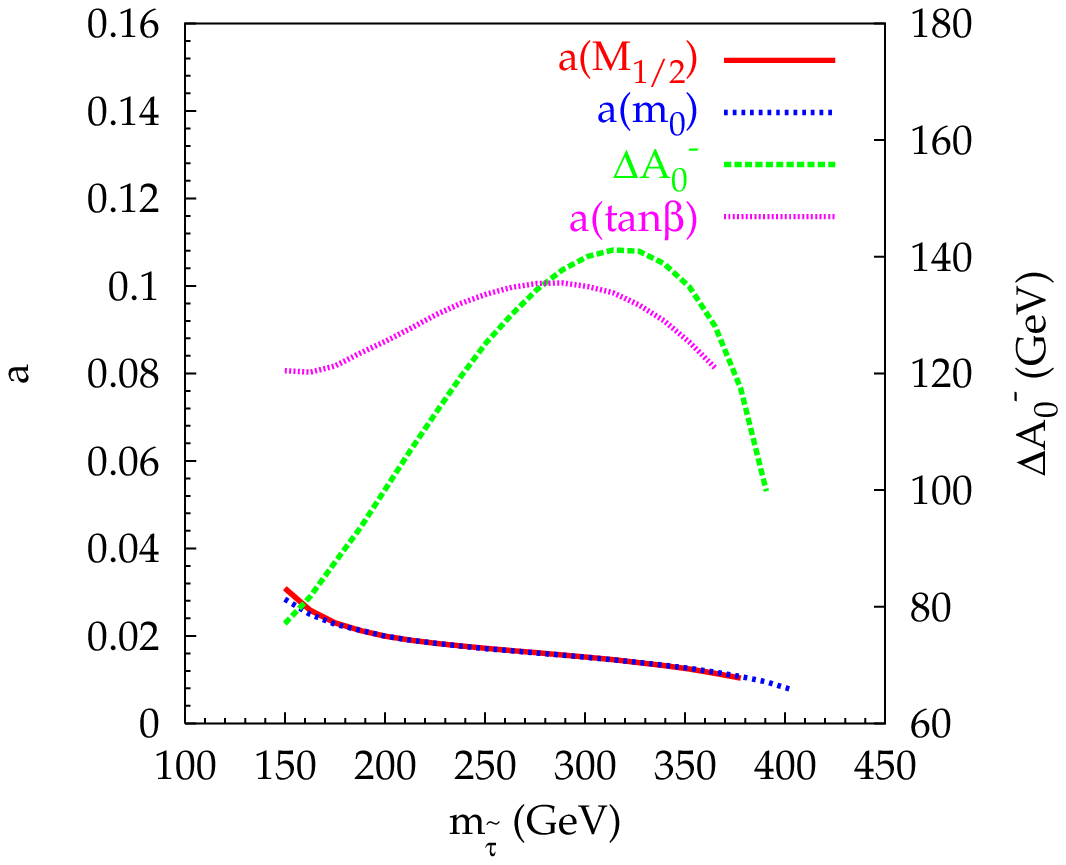, width=\wth}}
\end{picture}
\caption{Required fractional accuracies in mSUGRA parameters along slope
  S1 in order to achieve WMAP precision. The range of the abscissa
  corresponds to $\mhf=350-920$ GeV. The $A_0$ sensitivity is given in absolute terms, $\Delta A_0$,
  since the reference point is $A_0=0$.}
  \label{fig:coanSugra}}

In the mSUGRA scenario, we assume that combined fits from the LHC
data agree with the mSUGRA predictions, and that $m_0, M_{1/2}$
and $A_0$ can be  constrained from observables involving other
particles than just those relevant for the co-annihilation region.
If the theoretical predictions are all under control, this means
that  we could predict the relic density without having accurate
information on the stau mass or on $\Delta M$ (which might be
quite difficult if not impossible). How small should then be the
precision on the experimentally reconstructed GUT scale parameters
to match the $10\%$ precision of WMAP? Fig.~\ref{fig:coanSugra}
shows the required accuracy, on both $m_0$ and $\mhf$, to be from
around $1-3\%$, with more accurate measurements being required for
heavier staus. This is not a surprise since $\mhf$ is directly
related to the LSP mass and $m_0$ contributes directly to
$\mstauo$ and therefore small changes may make co-annihilation not
work. More accuracy is needed for higher mass scales since the
mass difference is very small there and the usual $\neuto \neuto$
annihilation is negligible. This is in line with the observation
that the required precision is essentially set by the Boltzman
factor, $\exp{-\Delta M/T_f}$, with $T_f \sim \mneuto/25$, as we
will discuss later. This also explains why the accuracies on $m_0$
and $\mhf$ almost coincide. It rests to see whether such an
accuracy can be reached considering that from previous analyses
mostly done in the bulk region, the precision ranged from $1\%$ to
$10\%$ depending on the point in parameter space, see for
example\cite{Armstrong:1994it}. Another LHC
analysis\cite{Polesello:2004qy}, also done in the bulk region,
showed that the precision on the extracted mSUGRA parameters was
such that a precision of about $3\%$ on the relic density can be
achieved. Although we believe that, because of the decay chain
$\tilde{q}\ra \neutt \ra
 \ser \ra \neuto$ present also in this scenario, some of these analyses can be
carried over to the co-annihilation region, the small $\Delta M$
value means that some of the accuracies may only be reached near
the onset of co-annihilation. We also see that $A_0$ should be
measured  within about $100$GeV. The accuracy on $A_0$ should be
met in analyses similar to those carried for the bulk region,
where an accuracy of the same order as we require is
found\cite{Polesello:2004qy}. The accuracy on $\tgb$ is about
$9\%$ and is again about what is found for a particular bulk
region point in a LHC analysis\cite{Polesello:2004qy}.
Sensitivities to the standard model input parameters $m_b$,
$\alpha_s$ and $m_t$ are all negligible here.

\subsection{Accuracies on the relevant physical parameters}
The relevant physical parameters here are the masses of all the
light sleptons, in particular the $\stauo$, as well the mass of
the neutralino LSP. Once the weak scale parameters are set, we
take  as input parameters for the scalars, $\mstauo,\mstaut$ and
the mixing angle of the stau $\cos \theta_{\stau}$, as well as the
masses of $\ser$ and $\smur$.  For the neutralino one needs not
only the masses but also their couplings. These  involve the
diagonalising matrix. To be consistent we always start from the
set $M_1,M_2,\mu$ and $\tgb$, although as we discussed earlier the
co-annihilation region will be most sensitive to $M_1$ through
$\mneuto$. $\Delta M$ is a crucial parameter and we analyse its
accuracy through small changes in the physical mass $\mstauo$. All
other parameters are then fixed. \FIGURE{
 \unitlength=1.1in
\begin{picture}(4,2.6)
\put(0.8,0){\epsfig{file=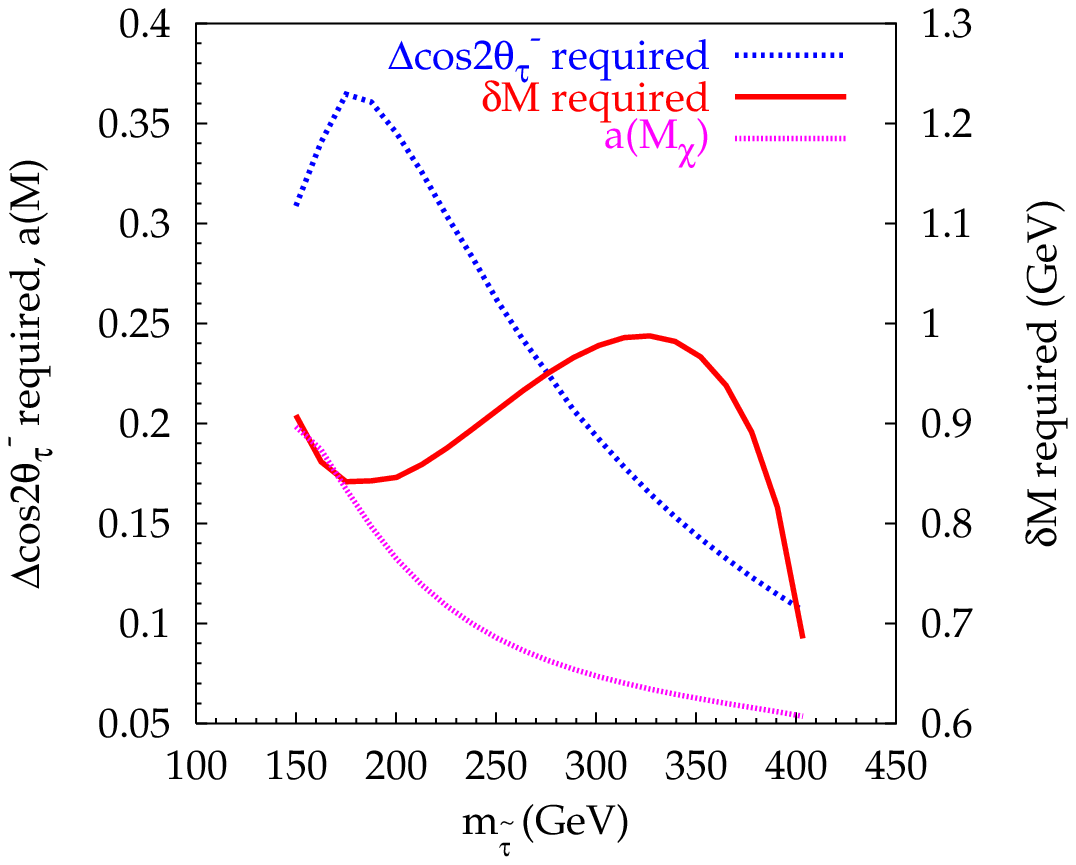,
width=\wth}}
\end{picture}
\caption{ Required accuracy on the stau-neutralino mass difference
obtained by varying the $\stauo$ mass while keeping all other
parameters fixed. Accuracy on $\cos 2\theta_{\stau}$ is also
shown. The required accuracy on the mass of $\neuto$ is performed
by keeping $\Delta M$ constant.  $\delta M=\delta(\Delta M)$. The
abscissa range corresponds to $\mhf=350-920$ GeV.}
  \label{fig:pud2}}
\FIGURE{ \twographsg{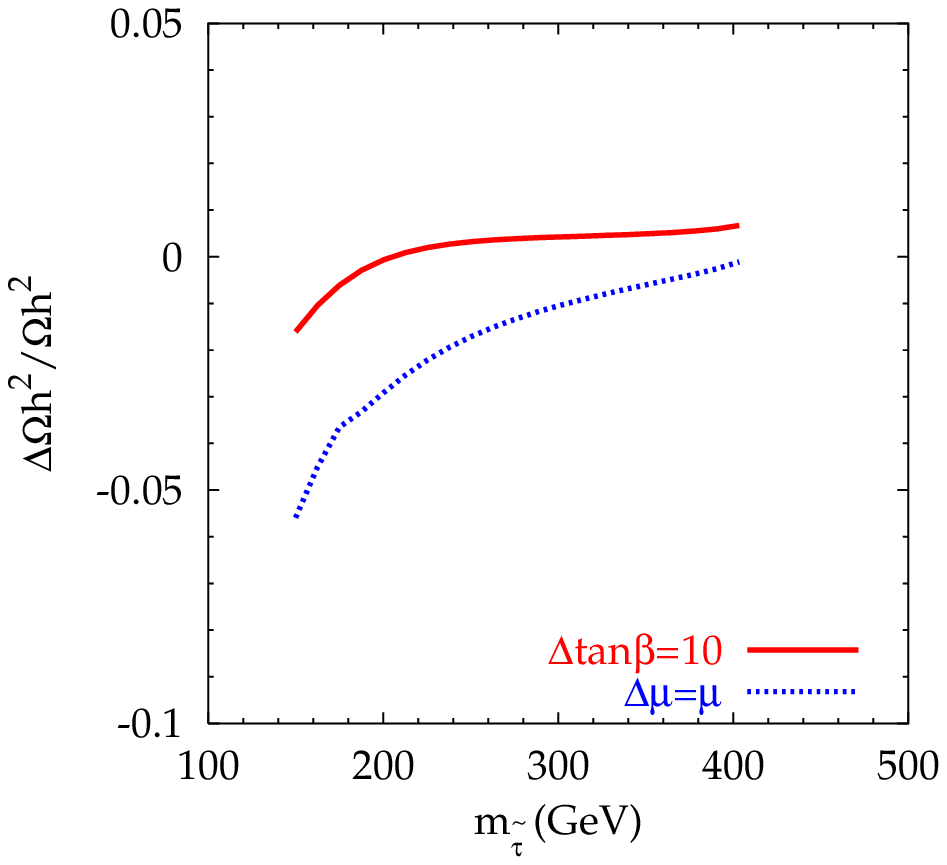}{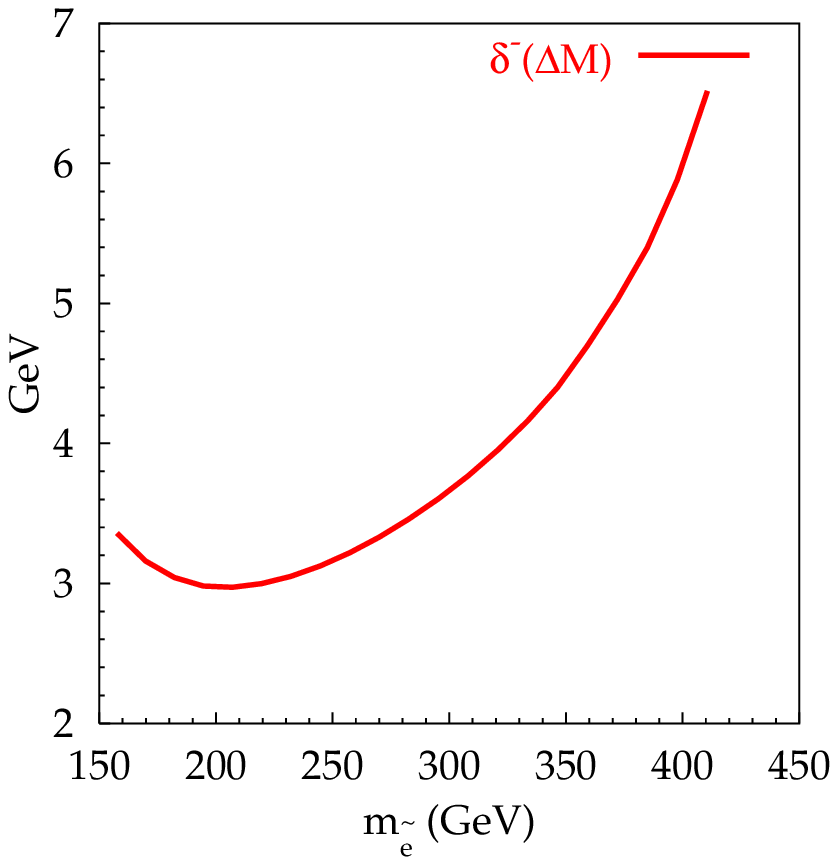}
\caption{(a) Keeping $\Delta M$ constant we give the fractional
variation in the relic density due to a $100\%$ change in $\mu$
and $\tgb$. (b) Accuracy on the common (right)selectron and smuon
masses given in terms of $\delta^-(\Delta M)=
\delta(\mser-\mneuto)$ in order to achieve WMAP precision.
\label{coan-massconst-msmu}}}

Fig.~\ref{fig:pud2} shows that the mass difference must be
measured within just less than 1GeV. Qualitatively this can be
understood from the fact that changing $\mstauo$ affects those
channels where co-annihilation involving $\stauo$ is important and
crucially depend on the Boltzman factor. Thus an approximate
derivation of the accuracy on the relic density can be arrived at
by looking at the variation of the Boltzman factor $d \ln \Omega
h^2 \simeq \delta \Delta M/T_f \sim 25 \delta \Delta M/\mneuto
\sim 25 \delta \Delta M/\mstauo$ with small variations caused by
the weight of the other channels. This accuracy can be turned into
an accuracy on $\mstauo$, $a(\mstauo)\sim 1/\mstauo(GeV)$, leading
to an accuracy that ranges from $0.7\%$ for the low $\mstauo\sim
150$GeV to $0.25\%$ for $\mstauo \sim 400$GeV.  $\cos 2
\theta_{\stau}$ must be measured to within $0.35$ (when its value
is about $\sim 0.92$) to about $0.1$ (when its value is almost
$1$). The reason it is larger at low $\mstauo$ is  that the stau
dependence is almost all contained in $\stauo \neuto \ra \tau
\gamma$, whereas as $\Delta M$ increases, more stau channels are
involved, for example $\stauo \stauo$ annihilation through
$Z$-exchange. We now seek the accuracy on the overall scale,
defined as the accuracy that is required on $\mneuto$ but leaving
$\Delta M$ unchanged. This of course means that we also need to
change all slepton masses by the same change that occurs in
$\mneuto$. The purpose of the exercise is to determine whether all
of  the sensitivity on the masses comes from $\Delta M$. Moreover
it may well be that measuring individual masses, especially the
LSP mass, may be more difficult in some situations than measuring
the mass differences. To determine this accuracy, we have chosen
to vary the $M_1$ parameter, leaving all other parameters of the
neutralino sector intact. This means however that small changes in
both the neutralino masses (notably the LSP) and of the couplings
of the LSP will be affected. The same change in the LSP mass is
then applied to all the lightest sleptons such that all relevant
mass differences and in particular $\Delta M$, remain unchanged.
We find an accuracy, corresponding to the $10\%$ WMAP precision,
ranging from $20\%$ for the low $\stauo$ mass to $5\%$ for
$\stauo$ masses around $400$GeV. This is consistent with a simple
scaling law for the relic density through the scaling of the cross
sections. At the higher end of the spectrum, $\Delta M$ is
negligible and there is essentially only one mass scale set by the
neutralino mass. All participating cross sections will scale as
$1/M^2$, once we freeze the Boltzman factor. Then the required
uncertainty on $M=\mneuto$ is half the uncertainty on the
measurement of the relic density. For lower $\stauo$ masses, terms
of order $\Delta M/M$ give non negligible effects. Note that this
accuracy is about a factor $20$ less precise than the one we
obtained by letting $\Delta M$ vary. Therefore the priority will
be on a very precise measurement of $\Delta M$ although precision
measurement of the overall scale is also demanded. We can also
ask, if $\Delta M$ is measured very precisely do we also need to
measure $\tgb$ and $\mu$, both of which enter the couplings of the
sleptons to the LSP? To find any appreciable change in this
context requires varying these two parameters quite substantially
while fixing $\Delta M$. With a $100\%$ change in $\tgb$, the
effect on the relic is below the $2\%$ level,  getting smaller at
higher masses as shown in Fig.~\ref{coan-massconst-msmu}. Doing
the same with $\mu$ we obtain a 5\%  change at low $\mstauo$ which
gets rapidly
smaller as we get to higher masses.\\
 Making sure that we are in
the co-annihilation region requires in addition the measurement of
the smuon and selectron masses. Fig.~\ref{coan-massconst-msmu}b
shows that we need to measure these to about $1.5\%$ to achieve
WMAP precision. This is obtained by varying the selectron and
smuon mass by the same amount, keeping everything else unchanged.
Very recent simulations\cite{Bambade:2004tq,Martyn:2004jc} of the
co-annihilation region, done in the context of a linear collider,
indicate that the relevant parameters $\Delta M$, $\mneuto,
\mstauo$ and $\mser, \msmur$ can all be measured with the required
accuracy to meet WMAP precision if one is in a region with large
enough $\Delta M$ (above $5GeV$ say). For increasingly smaller
$\Delta M<5$GeV associated with higher $\stauo$ masses, the
situation is more problematic and will need centre-of-mass
energies in excess of $600$GeV. Future investigations need also to
study precisely whether the accuracy on the measurement of the
mixing angle can be achieved in such a scenario. Detailed
simulations\cite{Boos:2003vf} at the LC have been performed but only in
a scenario with quite large $\Delta M\sim 80$GeV and $\stauo$
masses around $150$GeV, the lower end of slope S1. One can obtain
for example $\cos 2\theta_{\stau}=0.987\pm 0.06$.

\FIGURE{
 \unitlength=1.1in
\begin{picture}(4,2.6)
\put(0.8,0){\epsfig{file=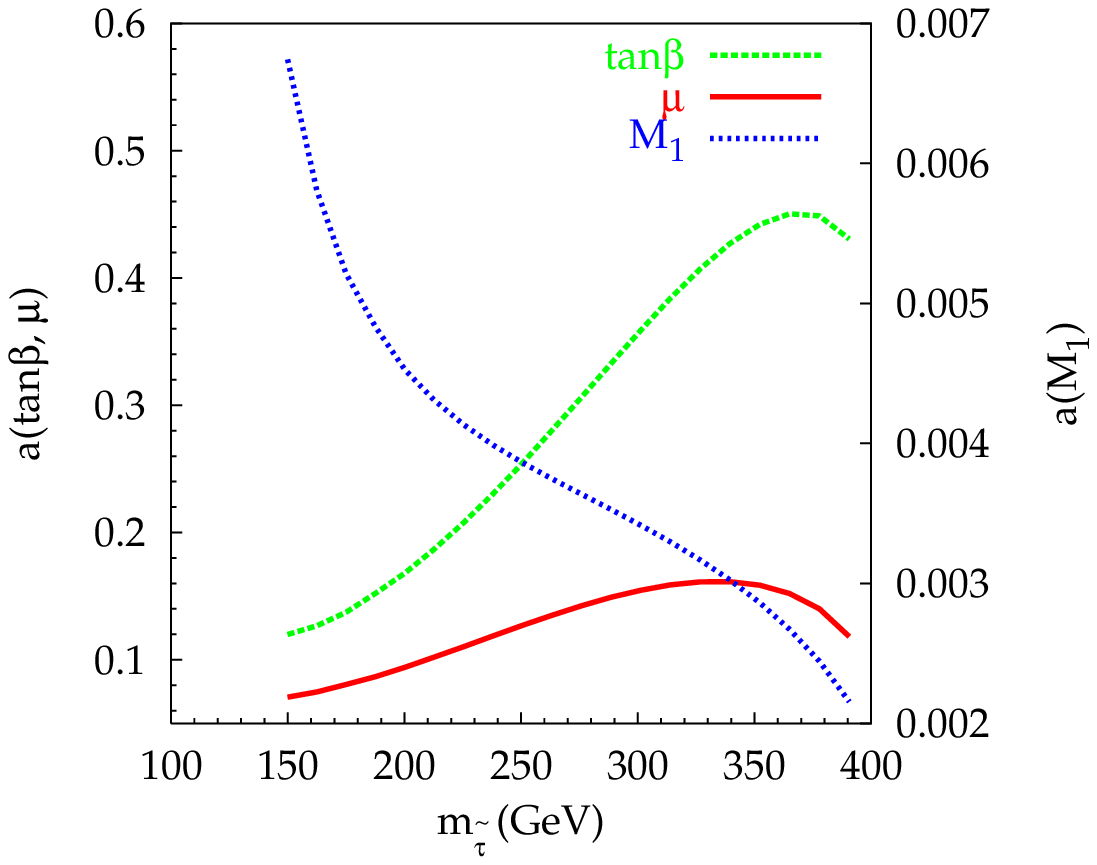, width=\wth}}
\end{picture}
\caption{Required experimental information along slope S1 in the
PmSUGRA
  scenario in order to achieve WMAP accuracy on the predicted relic
  density.
  Fractional
  accuracy required on $\mu$, $M_1$. The abscissa range corresponds to $\mhf=350-920$
GeV.}
  \label{fig:pud1}}

We could also ask what precisions on $\mu, \tgb$ and $M_1$ are
needed if we do not constrain $\Delta M$ to a fixed value. For
$M_1$, we see in Fig.~\ref{fig:pud1} that we require a few per-mil  accuracy. This  is the accuracy that we found for
$\mstauo$. The latter  can be derived trivially from $\Delta M$ in
Fig.~\ref{coan-massconst-msmu}. This is another indication that
most of the accuracy is set by $\Delta M$. However if one tries to
derive the accuracy on $\mu \tgb$ without constraining the
physical masses, by looking at variations on the Lagrangian
parameters, Fig.~\ref{fig:pud1} shows that  one might also need to
measure these two parameters quite precisely to arrive at a
precise determination of the relic density. This shows, when
compared to our previous analysis, that most of the $\tgb$ and
$\mu$ accuracy is contained in $\Delta M$. In fact, the $\tgb$
accuracy shown in Fig.~\ref{fig:pud1} can be derived from the
approximate formulae given in Eqs.~\ref{eq:mneutco},\ref{mstauo-1}
and \ref{mstauo-2}. Note in passing that the accuracy on $M_1$ is
more demanding than that for $\mhf$ in the pure mSUGRA analysis,
although we might think at first that since $M_1$ is directly
proportional to $\mhf$, we should expect similar accuracies. It
should be remembered though that $\mhf$ contributes quite
substantially to $\mstauo$ through the RGE running in this
scenario, see Eq.~\ref{mstauo-1}. A large cancellation  in $\Delta
M$ results which explains the weaker dependence on $\mhf$  in
mSUGRA.

\section{\label{sec:largetb}Higgs Funnel}
\subsection{Characteristics of the funnel region}
Rapid and efficient annihilation can occur through the Higgs
resonance. Because of the Majorana nature of the neutralino,
resonant enhancement is only obtained via the pseudoscalar Higgs
boson. The heavier CP-even Higgs which may be of equivalent mass
is completely buried because of the P-wave suppression. In this
situation $A\ra b \bar b$ is by far the dominant channel at high
$\tgb$, with some contribution from $\tau \bar \tau$. We have
checked that for the funnel region we are studying these two
contributions accounted for more than $98\%$ of the relic density,
the rest is annihilation into other fermion pairs. This also
confirms that annihilation through $Z$ exchange or sfermion
exchange is not significant. Therefore we should keep in mind that
what sets the stage here is the quantity $2 M_{\neuto}-M_A$ and
the width of the pseudoscalar since they define the $A$ resonance
profile. A precise prediction of the relic density also relies on
a precise determination of the ${\neuto \neuto A}$ coupling and
also the $Ab\bar b$ (and to some extent $A \tau \bar \tau$). In
fact adding the contributions of the $b$ and $\tau$ final states
amounts, to a very good approximation, to expressing these two
contributions in terms of the total width of the pseudoscalar. To
track down the dependence of $\Gamma_A$ remember that $\Gamma_A
\propto M_A \tgb^2 (m_b^2+m_\tau^2)$. The $m_b$ and $\tgb$
dependence will be made clearer later. As
known\cite{Griest:1990kh} the peak structure in the cross section
gets smeared once we perform thermal averaging in order to arrive
at the relic density. Approximations based on an expansion in the
relative velocity only grossly reflect the result. Keeping this
important fact in mind, we shall nonetheless, in order to see
which most important parameters are at stake, set the relative
velocity of the neutralino to zero\footnote{In the situation we
will be studying, this approximation should not be too
untrustworthy since we will be somewhat away from the pole.
As  we will see $(M_A-2\mneuto)/\Gamma_A$ ranges from 1.4 to 10.
Moreover, the width at $\tgb=50$ is not that small. Points
directly on the resonance lead  to annihilation rates that are too
low to be within the WMAP range.} . The important parameters are
then encoded in
\beqn
\label{Aprofile} <\sigma v>_{v=0}^{-1} \propto
\frac{\left((2\mneuto)^2-M_A^2\right)^2+\Gamma_A^2 M_A^2}{\mneuto
\;
\hat{\Gamma}_A\; g_{\neuto \neuto A}^2} \sim \frac{4 \mneuto
\Gamma_A}{g_{\neuto \neuto A}^2}
\left(4\left(\frac{M_A-2\mneuto}{\Gamma_A}\right)^2+1\right).
\nonumber \\
\eeqn

\noi $\Gamma_A$ is the total width of the pseudoscalar Higgs while
$\hat{\Gamma}_A$ is its total width into non-supersymmetric
particles. In our scenario the difference is only due to the
contribution of $A\ra \neuto
\neuto$, which is negligible as it only occurs very near threshold
and in any case does not compete with $A\ra b \bar b$
for such high $\tgb$ values. We have checked that it constitutes a
mere $0.05\%$ of the total width and could thus be safely
neglected. By default, this SUSY contribution
is not added in {\tt micrOMEGAs}. When looking for the accuracy on the total
width in
PmSUGRA, we will in fact vary the $A b \bar b$ vertex through a
change in $m_b$ so that in effect we are identifying
$\Gamma_A=\hat{\Gamma}_A$. In the same way, an $\alpha_s$
variation in PmSUGRA will affect both $\Gamma_A$ and
$\hat{\Gamma}_A$. As we will see, the Higgs funnel occurs for
rather large values of $\mu$ and relatively large $M_A$. In mSUGRA
these two parameters are strongly related. The neutralino is still
essentially of a bino nature, but in order to provide a coupling to the $A$
some Higgsino mixing is needed. The coupling ${\neuto
\neuto A}$ is, approximately, controlled by
\beqn
\tilde{g}_{\neuto \neuto A} \propto
\frac{M_Z}{M_1^2-\mu^2} (M_1 s_{2\beta}+\mu )\sim
-\frac{M_Z}{\mu} \sim  -\frac{M_Z}{\mneutth},
\eeqn
where the second step of the approximation assumes large $\tgb$
and $M_1
\ll
\mu$.

\noi $\Gamma_A$ is an important parameter. Its main contribution
is from  $A\ra b \bar b$ which is subject to large QCD
corrections. The latter can be absorbed by the use of an effective
$b$ quark mass $m_b^{eff}$ expressed in terms of the
$\overline{MS}$ $b$ mass at the scale of the Higgs boson and some
subleading $\alpha_s$ corrections. The Higgs coupling also
contains SUSY corrections that affect the bottom Yukawa
coupling. These can be resummed and we parameterise them through
the quantity $\Delta m_b$. In mSUGRA $\Delta m_b$ is also an
important parameter for the weak scale boundary condition at high
$\tgb$. It is derived in {\tt SOFTSUSY}~from
\begin{equation}
m_b(M_Z)^{\overline{DR}}_{MSSM}=m_b(M_Z)^{\overline{DR}}_{SM} / (1
+\Delta \mb), \label{deltab}
\end{equation}
which re-sums some $\sim {\mathcal O }(1)$ corrections in the SUSY
loop contributions $\Delta \mb$.

\noi An effective Lagrangian for the couplings of the Higgses can be
written as
\begin{eqnarray}
\label{leffhbb}
{\cal L}_{eff}= \sqrt{4\pi\alpha}_{QED}
\frac{m_b^{{\rm eff}}}{1+\dMb}\frac{1}{2 \mw\sw}
\left[ - H b\bar{b}\frac{\cos\alpha}{\cos\beta}
\left(1+\frac{\dMb \tan\alpha}{\tan\beta}\right)\right.\nonumber\\
+ i \gamma_5 Ab\bar{b}
 \tan\beta\left(1-\frac{\dMb}{\tan\beta^2}\right)
+h b\bar{b}\left.
\frac{\sin\alpha}{\cos\beta}
\left (1-\frac{\dMb}{\tan\alpha\tan\beta}\right)\right],
 \end{eqnarray}
where the angle $\alpha$ originates from the diagonalisation of
the CP-even Higgs system.

 $\Delta \mb$ in the funnel region we are
studying is about $25\%$. Its theoretical determination requires
the knowledge of a large part of the SUSY spectrum, especially 
the third generation squarks, gluinos and also the Higgsino. All
of these may prove difficult to measure as these masses are high.
One should therefore try to determine $\Gamma_A$ directly from
experiment. As an alternative we may also wonder whether $\Delta
\mb$ could be extracted from precision measurements of the light
Higgs, $h$. We feel that this will be extremely difficult if one
is in the decoupling regime that occurs for large $M_A>200GeV$.
In the decoupling regime, $h$ is essentially SM-like and the
$\Delta \mb$ correction is screened being suppressed by the
factor $M_Z^2/M_A^2$ (although large $\tgb$ delays this decoupling
in $h b \bar b$, see \cite{Carena:2001bg}). To wit, normalised
to the SM coupling, the $h$ and $A$ couplings write respectively,
as
\beqn
\label{eq:h-delm}
h_{bb} \simeq \sin(\beta-\alpha) - \cos(\alpha-\beta) \tilde{A}_{bb}, \nonumber \\
A_{bb}=\tilde{A}_{bb} \left(1-\frac{\dMb}{\tan\beta^2}\right)
\simeq \tilde{A}_{bb}; \quad \tilde{A}_{bb}=\frac{\tgb}{1+\dMb}.
\eeqn

\noi $\sin(\alpha-\beta)$ is almost $1$ in the decoupling regime
and could be determined from the measurement of the $ZZh$
coupling,  for example. It would be interesting to examine in such
a scenario whether one could combine the extraction of
$\tilde{A}_{bb}$  through a precision measurement of $h\ra b
\bar b$ at the linear collider and the LHC measurements of $M_A$ in order
to determine $A\ra b \bar b$ more precisely. There is also an
equivalent of $\Delta
\mb$ for $A\ra \tau \bar \tau$, however it is much smaller since
the corresponding Yukawa coupling is much smaller and  because it
does not feel the important QCD correction.

In our discussion of the Higgs funnel we will take  $\tan
\beta=50, A_0=0, \mu>0$. We parameterise the funnel region
through the slope S2, defined as a cubic, see Fig.~\ref{fig:funnelSpec}a\footnote{We found that a
quadratic was
  not sufficient to contain $\Omega h^2$ within the 2$\sigma$ range.}:
\begin{equation}
\frac{m_0}{\gev} = \sum_{i=0}^{i=3} c_i \left(\frac{\mhf}{\gev} \right)^i,
\end{equation}
where $c_i=\{ 814.88, -2.20022, 3.30904 \times 10^{-3}, -1.05066
\times 10^{-6} \}$.

\FIGURE{ \twographsg{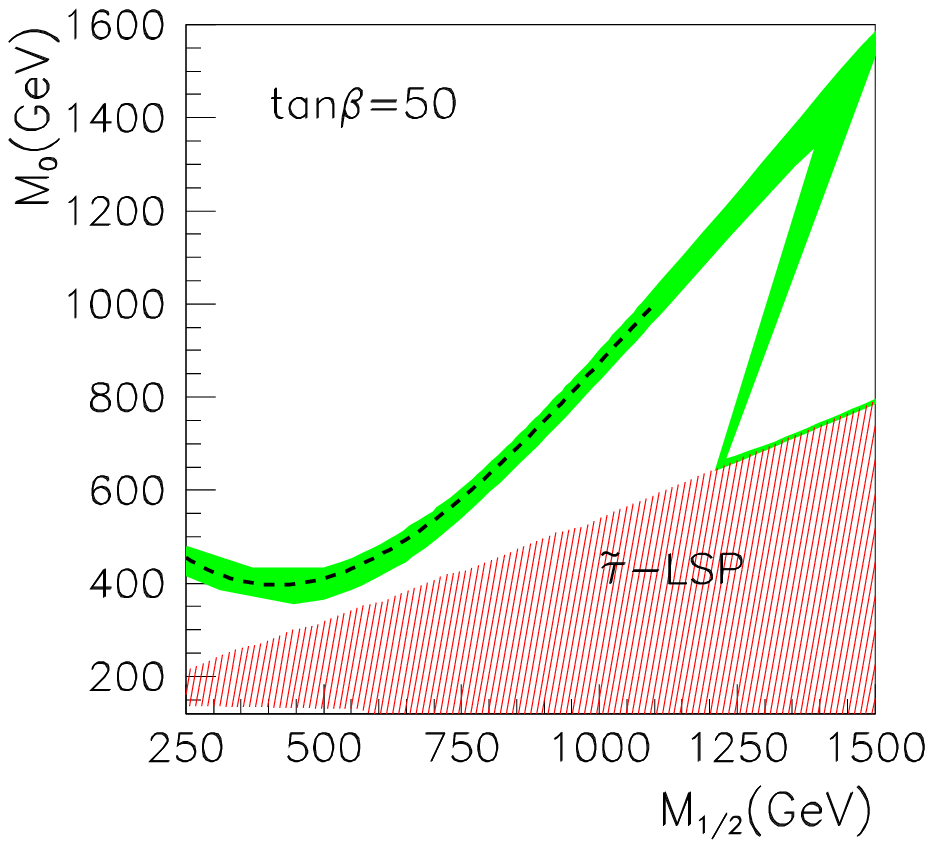}{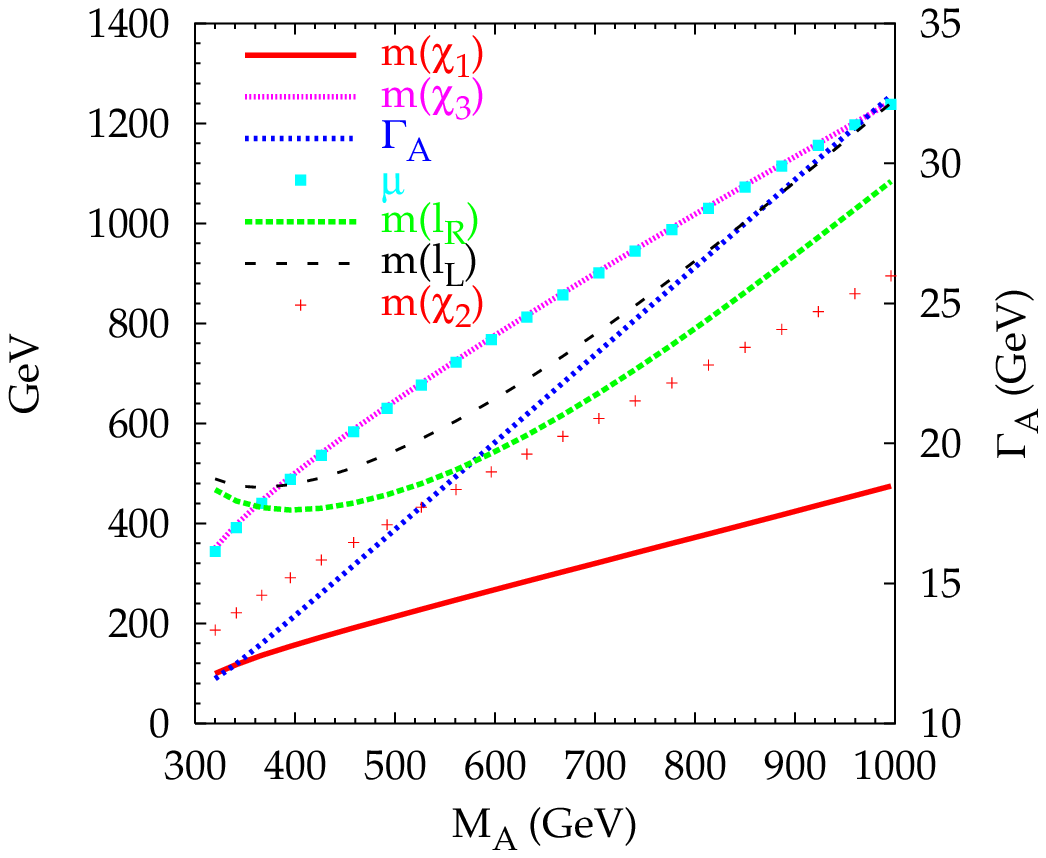}
\caption{a) Slope S2 in the $\m0-\mhf$ plane (dashed line) and the WMAP allowed region (green/light grey). The red area  is 
ruled out by the cosmological constraint that the LSP is neutral.  b) Mass spectrum of the relevant particles in the funnel
region. $\mhf$ is in the range $250$-$1100$GeV.}
  \label{fig:funnelSpec}}

%\FIGURE{
% \unitlength=1.1in
% \begin{picture}(4,2.6)
%\put(0.8,0){\epsfig{file=funnelSpec.eps, width=\wth}}
%\end{picture}
%\caption{Mass spectrum of the relevant particles in the funnel
%region. $\mhf$ is in the range $250$-$1100$GeV.}
%  \label{fig:funnelSpec}}

An important remark is that the funnel region has a rather  heavy
spectrum, see Fig.~\ref{fig:funnelSpec}. In particular the sfermions are very heavy making it
impossible to produce any sleptons at a $500$GeV machine. More
energy is needed for this. There is some chance for the associated
production of $\neutt$ (through $\neuto \neutt$ production), which
could serve as a good measurement of $\mneuto$ and indirectly of
the selectron mass. On the other hand, the $\gamma \gamma$ option
of a linear collider could bring important constraints on the mass
and couplings of the $A$ as well as its width. Combined with a
determination of $\mu$ from the LHC,  one could reconstruct the
parameter space that defines the funnel region through $\mneutth$ (observe
that $\mneutth$ tracks $\mu$ very well). The LHC can also
measure the mass of $A$, possibly up to $1$TeV, and might probe
$\Gamma_A$ through $A\ra \mu \bar \mu$. For future discussion it
is interesting to realise that $\Gamma_A/M_A$ here is about $3\%$,
slightly larger than the corresponding quantity for the $Z$
lineshape. Note also  that  $(M_A-2\mneuto)/\Gamma_A\sim 1.4-10$.

\subsection{Theoretical uncertainties in
mSUGRA} \FIGURE{\fourgraphs{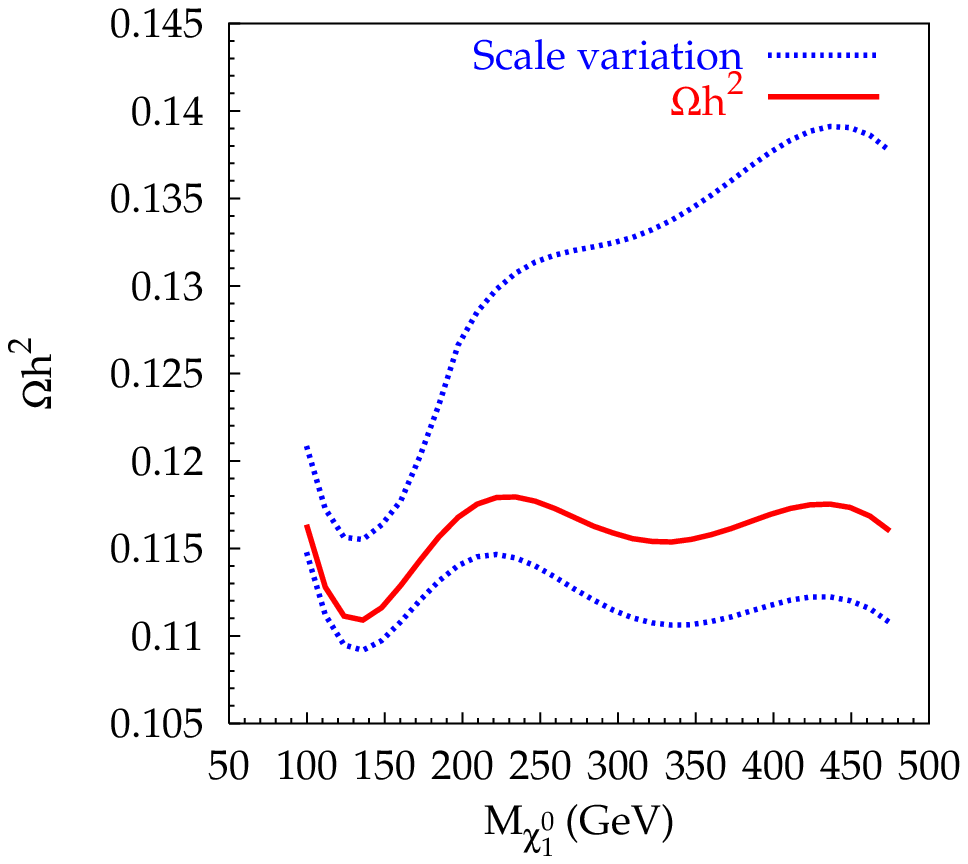}{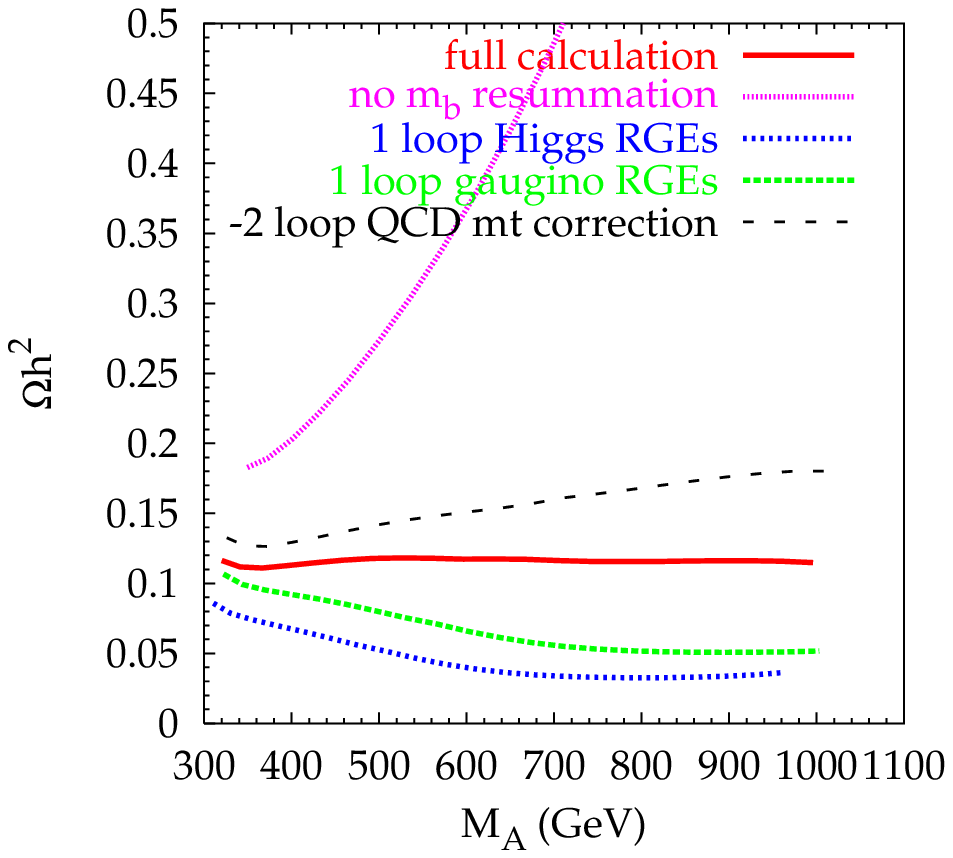}
{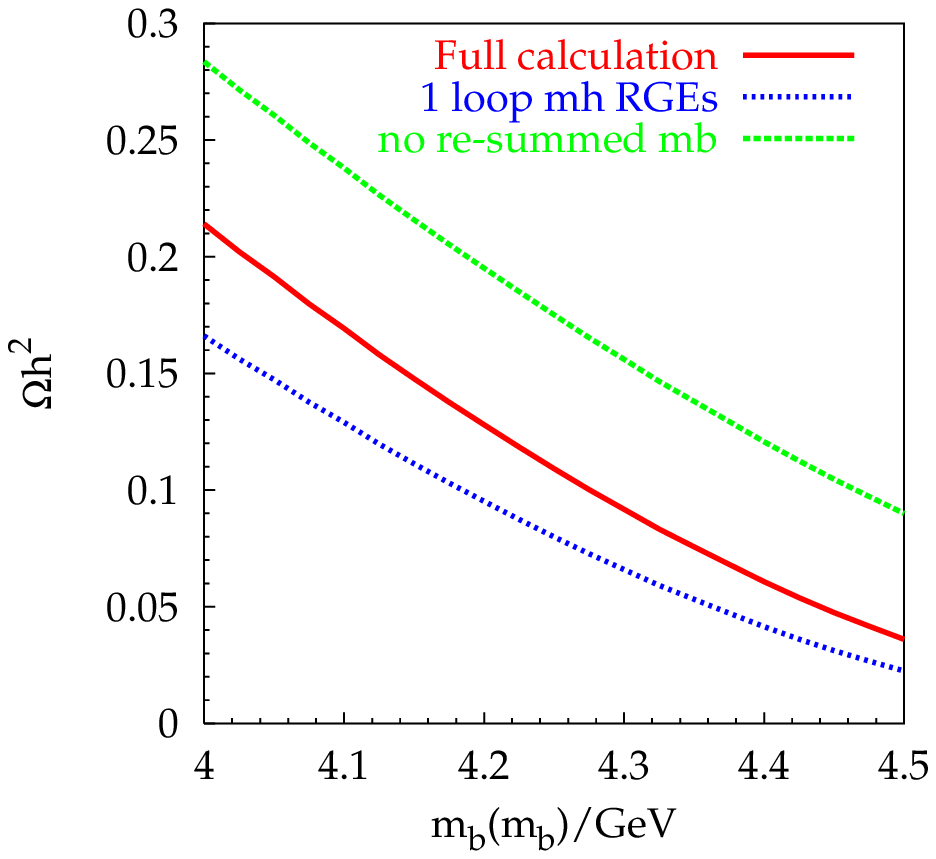}{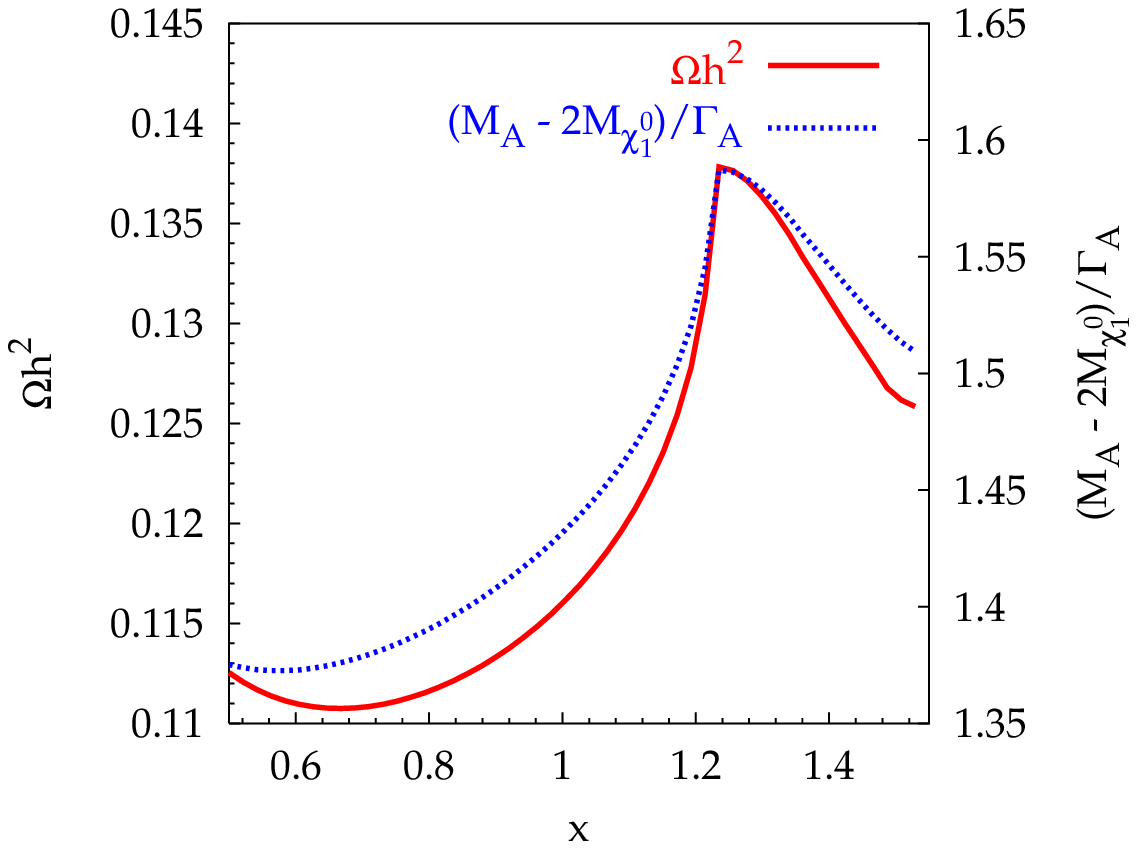} \caption{Relic density along slope S2. (a)
The central line gives the default
  {\tt
  SOFTSUSY} prediction, with the broken lines representing the limits due to
  scale uncertainty. The range of the abscissa corresponds to $\mhf=250-1100$
  GeV. $m_b(m_b)=4.23$GeV.
(b) Effect of different approximations in the RGE, see text for
details. (c) The dependence on the input value $m_b(m_b)$
including treatment of higher orders related to the $b$ mass (and
coupling Yukawa) for $\mhf=1100$GeV  on mSUGRA slope S2. The
meaning of the different curves is explained in the text.(d) Scale
dependence of the relic density compared to the scale dependence
of the relevant parameter controlling the dominant contribution of
the $A$ resonance for $\mhf = 1100$GeV GeV on slope S2. x is
defined in Eq.~\protect\ref{x}.}
  \label{fig:funSlope}}

The funnel region involves a heavy spectrum. Since it may
not be possible to have access to all the relevant parameters, one
may have to partially rely on some mSUGRA assumptions.  It is
therefore important to inquire about the theoretical uncertainties
in mSUGRA when predicting the relic density.

Fig.~\ref{fig:funSlope}a shows the scale dependence of $\Omega
h^2$ along the funnel slope S2 which, in passing, approximates the
WMAP data rather well. The scale dependence increases from $5\%$
at low values of $M_{\neuto}$ to 20$\%$ at higher values.
$M_{SUSY}'
\sim \{1.5-2\} M_{SUSY}$ did not yield physical points, possessing
unstable saddle-point electroweak vacua ($M_A^2(M_{SUSY}')<0$).
Such values of $M_{SUSY}'$ were therefore not included in the
scale variation. Fig.~\ref{fig:funSlope}b shows the dramatic
effects, especially for increasing $M_A$, of some higher order
loop contributions in the spectrum calculation for slope S2 and a fixed value
of
$m_b(m_b)$. Not re-summing the $\Delta m_b$ terms has a drastic
effect. Taking into account only the one-loop terms (instead of
two-loop terms in the default calculation) for the scalar Higgs
$m_{H_{1,2}}^2, m_3^2$ or gaugino masses in the RGE is not
sufficient at all. For $1$TeV pseudo-scalar mass, the 2-loop QCD
correction to the top mass is essential. Some of the effects of
the approximations can be counter-balanced by changing the
$m_b(m_b)$ input, Fig.~\ref{fig:funSlope}c.

To see if most of the dependence can be reduced if some physical
observables are fitted, we take the variable $(M_A - 2
M_{\neuto})/\Gamma_A$ that describes the $A$ resonance profile.
The illustration is made for a single point on S2: $m_0=1000$ GeV,
$\mhf=1100$ GeV (corresponding to the extreme right hand side of
Fig.~\ref{fig:funSlope}a). Fig.~\ref{fig:funSlope}d shows that, as
one varies the scale, there is a peak in the relic density at the
same location as in the  variable $(M_A-2
\mneuto)/\Gamma_A$ describing the $A$ resonance. This indicates
that a measurement of this ratio could help pin down the
uncertainties.

\subsection{Required accuracies on the mSUGRA parameters}

\FIGURE{ \twographsg{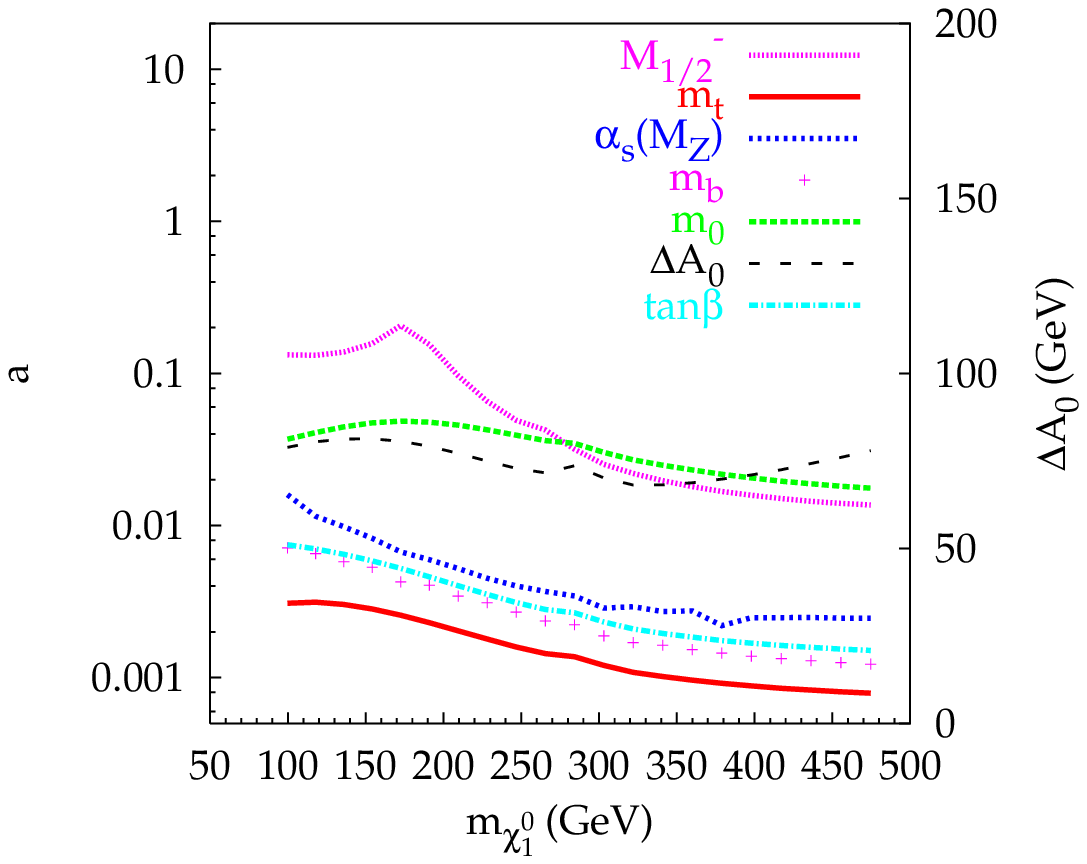}{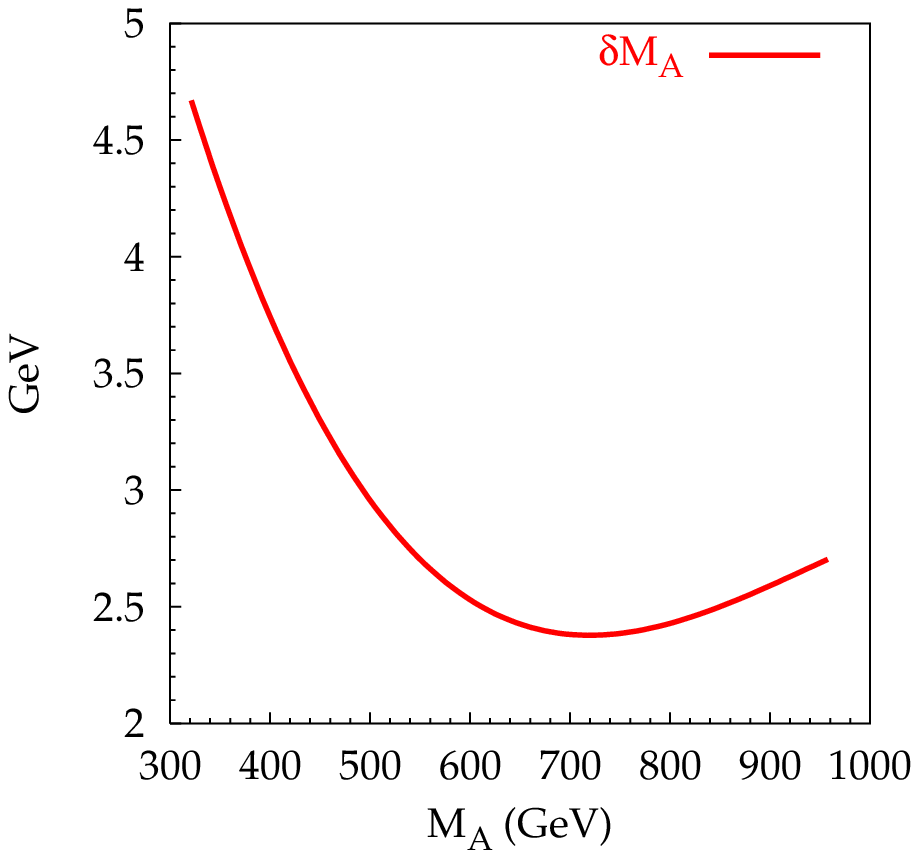}
\caption{(a) Accuracies $a$ required to achieve WMAP precision along slope S2
  in   the mSUGRA scenario. (b) Pseudoscalar Higgs width $\Gamma_A$ and
  required precision upon in $\delta \Gamma_A$ along slope S2 in the mSUGRA
  scenario.
The range of the abscissa corresponds to $\mhf=250-1100$
  GeV for both (a) and (b). This plot applies to the accuracy needed on $m_b$ when all other
  SM and mSUGRA input parameters are set (or known precisely). However we translate
  the $m_b(m_b)$ accuracy into the precision needed on $M_A$. }  \label{fig:funDerTh}}

In Fig.~\ref{fig:funDerTh}a, we see the accuracies required on
measurements
  assuming mSUGRA is correct. $m_t$ must be measured at the level of 0.6-0.2
  GeV, depending upon position along the slope. $\alpha_s(M_Z)$, $\mb$ and $\tgb$ must be known
at the per-mille level. Most of this dependence is from the
extraction of the $b$ Yukawa coupling. As we have seen this could
be controlled by measuring $M_A, \Gamma_A$. To match WMAP accuracy
a 2-5$\%$ measurement of $m_0$ is required, while  $M_{1/2}$ needs
to be known at the 1$\%$ level (although there is a
  region where an $\mathcal{O}(1)$ measurement of $M_{1/2}$ would suffice).
$A_0$ must be measured to 100 GeV accuracy. We have also attempted
to quantify how the accuracy needed on $m_b(m_b)$ from WMAP
translates into a precision on $M_A$. For this exercise we of
course still need to assume that all other SM and mSUGRA input
parameters are  known precisely so that a theoretical prediction
of $M_A$ within mSUGRA is possible. We find that one needs to
measure $M_A$ to less than $5$GeV. This would presumably require
the measurement of a line-shape by (for example) $A \rightarrow
\mu^+ \mu^-$~\cite{Armstrong:1994it}. If that were possible we
could then use $M_A$ as a trade-off for $m_b(m_b)$ in mSUGRA. On
the other hand we found that the width $\Gamma_A$ is insensitive
to $m_b(m_b)$ in mSUGRA. This should not be taken to mean that
with measurements of $\Gamma_A$ and $M_A$ {\em alone} we can predict the
relic density within WMAP precision.
%Indeed $m_b$ alone does not
%give the value of $M_A$ for example nor the other crucial
%parameters like the mass and couplings of the LSP.

\subsection{Pmsugra}
\FIGURE{\unitlength=1.1in
\begin{picture}(4,2.6)
\put(0,0){\epsfig{file=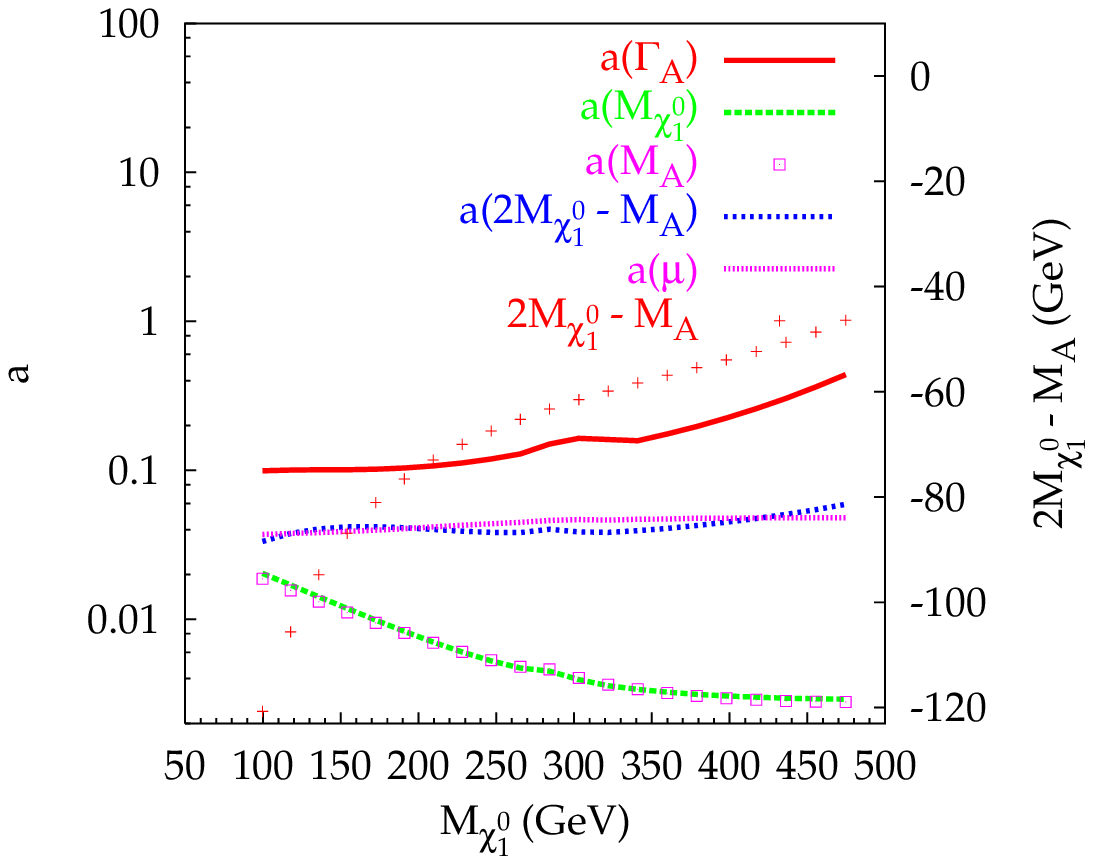, width=1.3\wth}}
\end{picture}
\caption{Accuracies $a$ required to achieve WMAP precision along slope S2
  in the PmSUGRA scenario. The range of the abscissa corresponds to
  $\mhf=250-1100$GeV.   $2\mneuto -M_A$ is also shown on the right hand axis.
}\label{fig:funExp}}

As we have already observed, the relic density in the funnel
region derives essentially from $\neuto \neuto \ra (b \bar b+ \tau
\bar \tau)$  through pseudo-scalar Higgs exchange. The important
physical parameters are $M_A, \Gamma_A, \mneuto$ as well as the
parameters that enter the $\neuto \neuto A$ coupling which are
mainly controlled by $\mu$, see Eq.~\ref{Aprofile}. For all
purposes $\mu$ can be equated with $\mneutth$, see
Fig.~\ref{fig:funnelSpec}. Let us therefore start by finding out
the accuracy needed on this physical parameter first while keeping
all other parameters fixed. To arrive at this accuracy we simply
change the value of $\mu$ at the Lagrangian level given at the
weak scale. Strictly speaking this also changes the value of the
LSP mass. In the situation we are in, $M_1, M_Z \ll \mu, \tgb \gg
1$, the induced change in $\mneuto$ will be a very small change as
can be seen from Eq.~\ref{eq:mneutco}. More important, a change of
$\mu$ will impact directly on the LSP coupling to the pseudoscalar
Higgs. Fig.~\ref{fig:funExp} corroborates this observation. We see
that one needs an accuracy of about $5\%$ on $\mu$ for a $10\%$
WMAP precision, which corresponds to the fact that $\Omega h^2
\propto \mu^2$. The evaluation of the accuracy on $\Gamma_A$,
independently of any of the other relevant parameters, is
triggered through a change in $\mb$. We find an accuracy on the
total width which is of the same order as the WMAP accuracy, in
accordance with Eq.~\ref{Aprofile}. We have also  looked at the
accuracy needed on $\alpha_s(M_Z)$, assuming one knows the other
parameters entering the calculation of $\Gamma_A$. We confirm that
the accuracy needed is proportional, and of the same order,  to
that found for $\Gamma_A$. The accuracy on the LSP mass is arrived
at by varying the  bino mass $M_1$, although this also very
slightly changes the $\neuto \neuto A$ coupling. We find an
accuracy on $\mneuto$ ranging from $2\%$ for $\mneuto \sim 100$GeV
to $0.2\%$ for $\mneuto \sim 450$GeV. This can also be translated
into an accuracy on the $A$ profile parameter $(2\mneuto -M_A)$
which we find to be, as expected, of the same order as $\mu$:
$5\%$ for WMAP accuracy. The accuracy of $M_A$ is derived by
varying $M_A$ but note that this also changes $\Gamma_A$,
$\Gamma_A \propto M_A$. As expected from Eq.~\ref{Aprofile} we get
an accuracy on $M_A$ which is practically the same as the one on
$\mneuto$, from $2\%$ to $0.2\%$ for WMAP precision.
Experimentally, this will be quite demanding, especially for the
largest of pseudo-scalar masses allowed in our funnel region. It
is important to add that once we choose the set $M_A, \Gamma_A,
\mneuto,\mu (\mneutth)$, the $\tgb$ dependence is very mild. To
confirm this we have tried to ``decouple" the $\tgb$ dependence in
the vertex $A f \bar f$ from the one in the coupling of the
neutralino. We do this by taking a point with the default value of
$\tgb$. We then change $\tgb$ and record the changes in the
neutralino mass matrix. This new neutralino mass matrix is then
substituted in the original default point, leaving all other
variables unchanged. We find for example that a $10\%$ change in
$\tgb$ does not affect the relic density beyond $1\%$. Although
negligible for WMAP accuracy, requiring PLANCK accuracy means that
a good determination of $\tgb$ might still be needed.

\section{\label{sec:focus}Focus point}
\subsection{The focus point landscape}

\FIGURE{\twographs{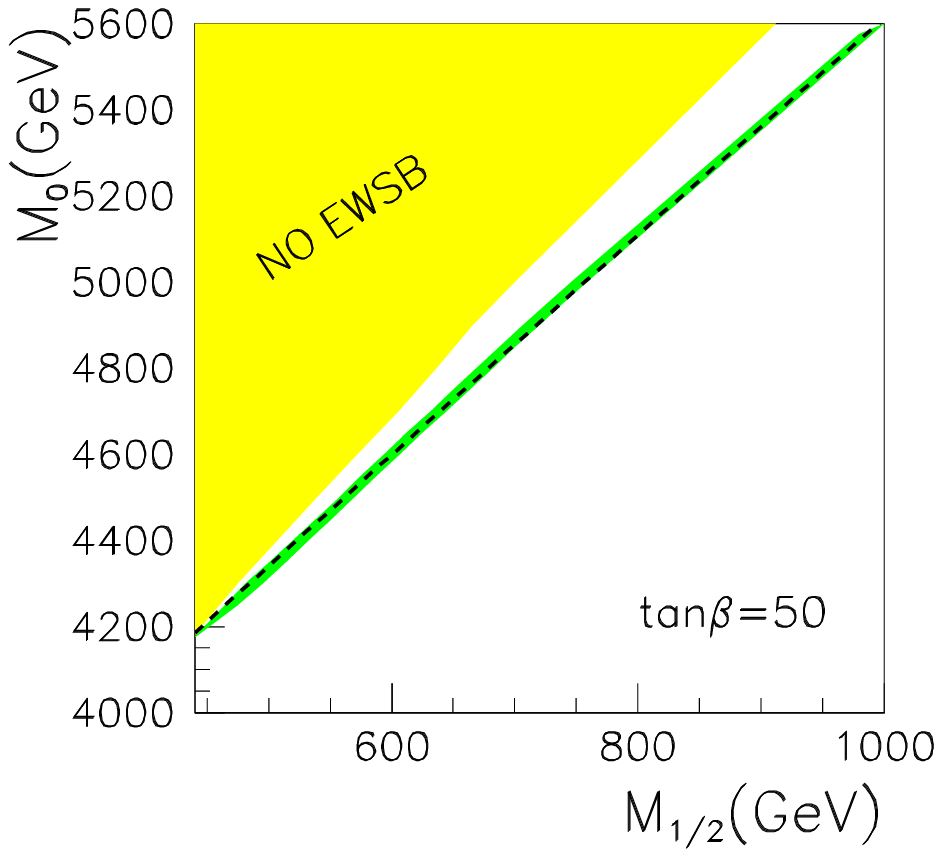}{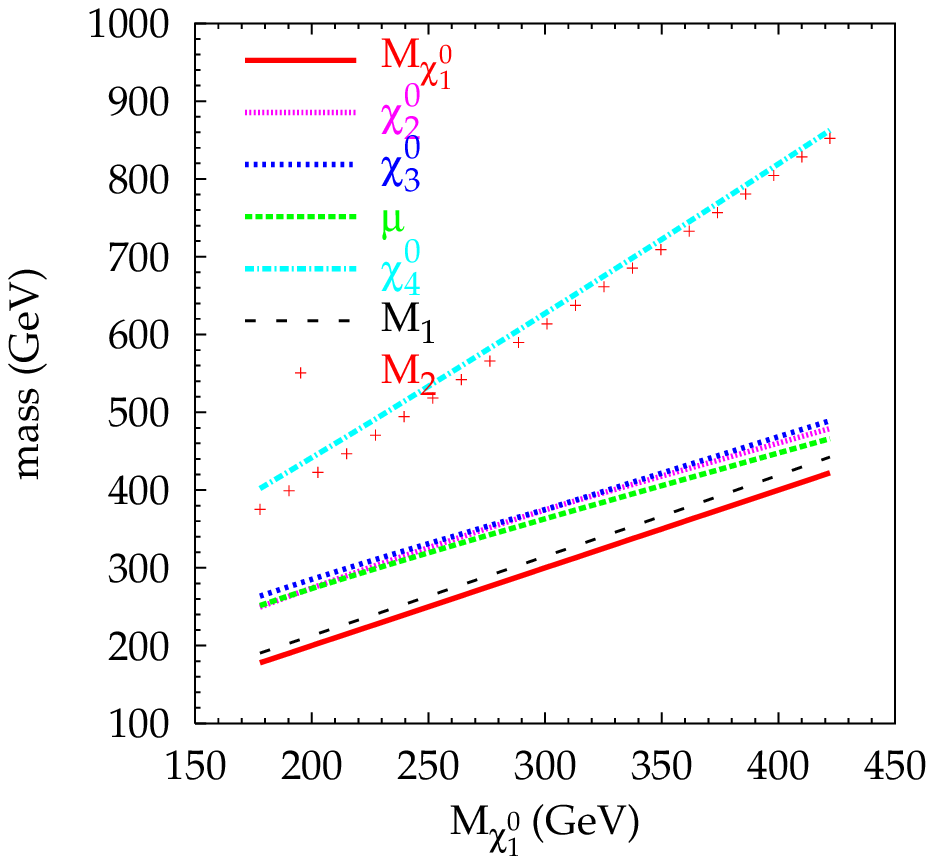}
\caption{a) WMAP allowed region in  the $\m0-\mhf$ plane (green/light grey) and slope S3 (dashed line). b) Mass spectrum in the {\tt SOFTSUSY} focus point region along slope S3. $\tgb=50$. }
\label{fig:focSpec}}
The  focus point region~\cite{Feng:1999mn,Feng:1999zg} corresponds
to high values of $m_0$ near the boundary of viable electroweak
symmetry breaking (see Fig.~\ref{fig:focSpec}a) where the value of $\mu$ drops rapidly. When
$\mu \sim M_1,M_2$,  the LSP has a significant Higgsino fraction,
furthermore the next-to-lightest sparticles ($\neutt$ or
$\chargopm$)  also have a significant Higgsino component and are
not much heavier than the LSP as shown in Fig.~\ref{fig:focSpec}b. Thus co-annihilation channels are
favoured. However, co-annihilation
should not be too efficient, otherwise the relic density is less
than what is measured. The currently acceptable focus point region
obtained with {\tt SOFTSUSY}, though requiring $\mu$ smaller than
in the co-annihilation and funnel regions, does not lead
$\mu<M_1,M_2$ but rather $M_1<\mu<M_2$. This still means that the
LSP has a sufficient Higgsino component, of about $25\%$, so that
stronger couplings to the $Z$ and $W$ take place, in particular through their
Goldstone component, and to some extent through Higgs bosons.

We will  examine a high $\tan \beta=50$ focus point compatible
with WMAP as obtained with {\tt SOFTSUSY} and take a slope ``S3''
in parameter space: $\tan \beta=50, \mu>0, A_0=0$ and
\begin{equation}
\frac{m_0}{\gev} = 3019.85 +  2.6928\frac{M_{1/2}}{\gev} -1.01648
\times 10^{-4} \left(\frac{M_{1/2}}{\gev}\right)^2.
\end{equation}

\noi with $\mhf$ in the range $440-1000$GeV, see Fig.~\ref{fig:focus-contrib}a. This means that all
sfermions are far too heavy (in excess of $4TeV$ or so) to be
accessible at any of the planned colliders. The interesting
feature of the spectrum is shown in Fig.~\ref{fig:focSpec}b. The
LSP mass ranges from about $150$ to $350$GeV with the
Higgsino-like neutralinos being some $100$ to $50$GeV heavier. Their
mass differences decrease for the higher masses and thus the
Higgsino content of the LSP increases. The smaller mass difference
suggests that co-annihilation may be more important for  higher
scales. The wino-like gauginos are above $400$GeV. From the point
of view of a linear collider,  an energy in excess of $800$GeV is
needed to unravel some properties of this scenario. Although not
shown on the plot, let us mention that the pseudo-scalar has a mass in
excess of $1$TeV and will most probably not be directly probed
at the LHC. The gluino mass is also above $1$TeV but,
unlike the pseudo-scalar Higgs, does not enter directly in the
prediction of the relic density.

\FIGURE{ \epsfig{file=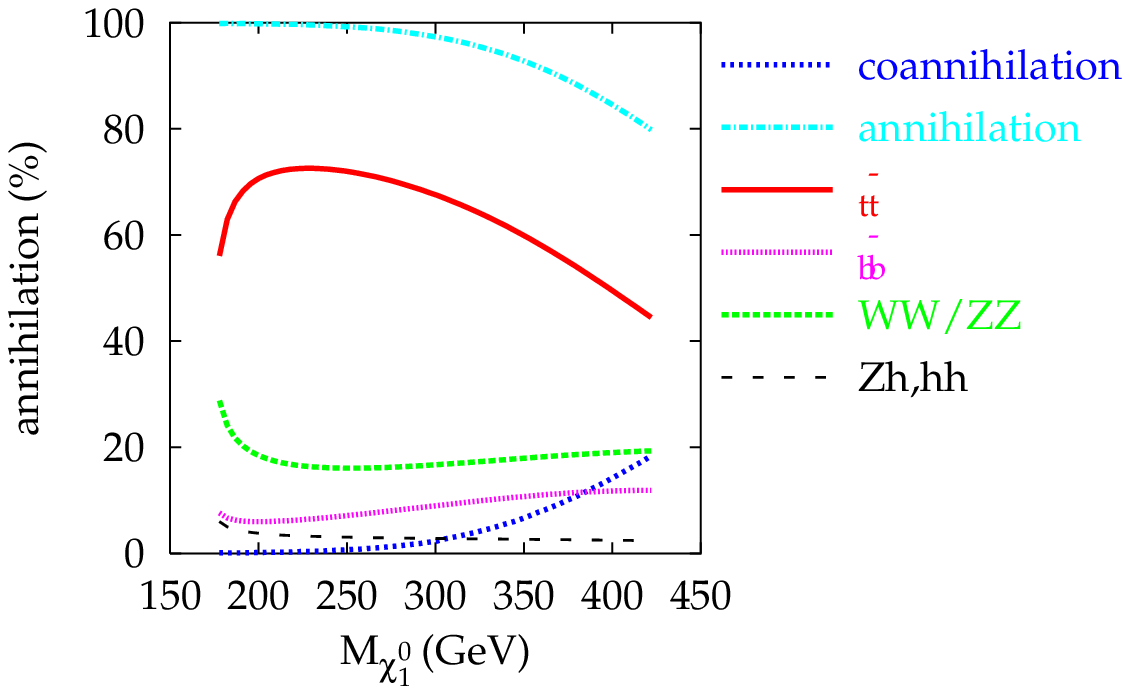,
width=1.5\wth}\caption{Contribution of the various channels in the
focus region. ``Annihilation" stands for all the channels which we
give individually as $t\bar t, b \bar b, W^+ W^-, ZZ, Zh$ and
$hh$. ``Co-annihilation"stands for $\neuto \neutt, \neuto
\neutth, \neuto
\chi_1^\pm$.
\label{fig:focus-contrib}}}

The contribution of the most important channels to the relic
density along slope S3 are displayed in
Fig.~\ref{fig:focus-contrib}. First of all, note that in our
scenario annihilation into tops is always the largest
contribution. This comes about through the Goldstone component of
$Z$ exchange, which picks up the large top Yukawa coupling.
Because of the relatively large Higgsino component, the neutral
Goldstone boson couples like $M_Z M_1/(M_1^2-\mu^2)$, with $\mu$
not much larger than $M_1$. We also note that we get a non
negligible contribution from annihilation into $b \bar b$.  This
contribution is due to $A$ exchange which, contrary to the $t \bar
t$ contribution via $Z$ exchange, gets a large $\tgb$ enhancement.
Properties of this contribution are similar to those in the funnel
region, though here it is diluted by comparison because of the
much larger pseudo-scalar masses involved and also because we are
far away from the pole. Although the $\neuto \neuto A$ coupling is
slightly larger than the $\neuto \neuto Z$ coupling, there is a
factor of $\mu/M_1$ between the two couplings, annihilation
through $A$ exchange is suppressed by the large mass of the Higgs
in the propagator, compared to the mass of the $Z$ for the $t \bar
t$ channel. Note the other annihilation channels into
$WW,ZZ,Zh,hh$ which account for $10\%$. For LSP masses around
$350$GeV, co-annihilation with the heavier neutralino  starts to
contribute since the $M_1,\mu$ mass difference gets smaller. From
these observations, the fact that quite a few channels and
mechanisms are contributing and that the masses involved are quite
high  will make a precise probe of this scenario rather difficult
for the colliders.

\subsection{Theoretical uncertainties}
Theoretically, the focus point regime is also fraught with difficulties.
The position of the focus point region is extremely sensitive to
the value of the $\overline{DR}$ {MSSM} top Yukawa
 coupling $h_t$~\cite{Allanach:2000ii,Allanach:2003jw}, which
 differs for different RGE codes~\cite{Allanach:2004jh}.
$h_t$ is fixed by the MSSM value of the running top quark mass. In
order to obtain this, supersymmetric and Standard Model
corrections are subtracted from the top pole
mass~\cite{Pierce:1996zz}. The way in which this is done
 differs in terms which are formally of higher order in the various codes. For
 example, some codes use pole masses for particles in the loops whereas some
 use running masses. The radiative corrections are sometimes calculated at
 different scales.
 For the case of {\tt SOFTSUSY}, values of the top quark mass
lower than the present 1$\sigma$ range are necessary to reach the
region where the relic density is in agreement with the WMAP
constraint if $\tan\beta$ is intermediate (e.g. 10-30). This is
the reason we concentrated on a high $\tan \beta=50$. We have
checked that the numerical results are similar (along a different
slope consistent with the WMAP measurement) for $\tan \beta=10$
and $m_t=172$ GeV.

\FIGURE{\twographs{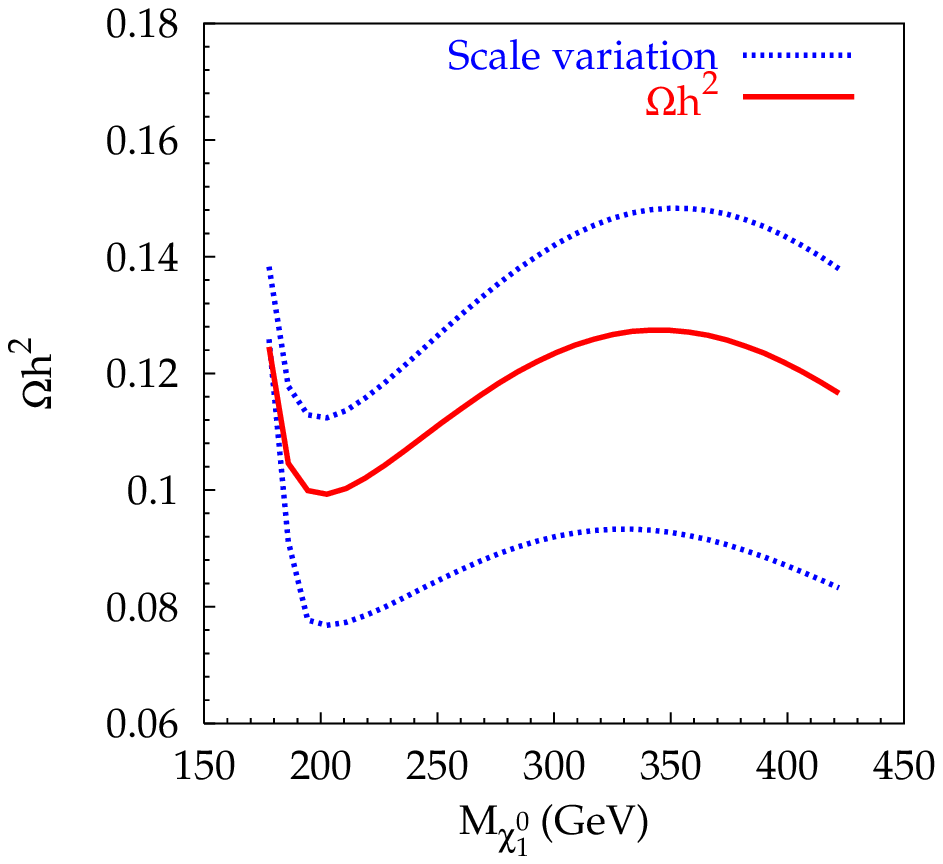}{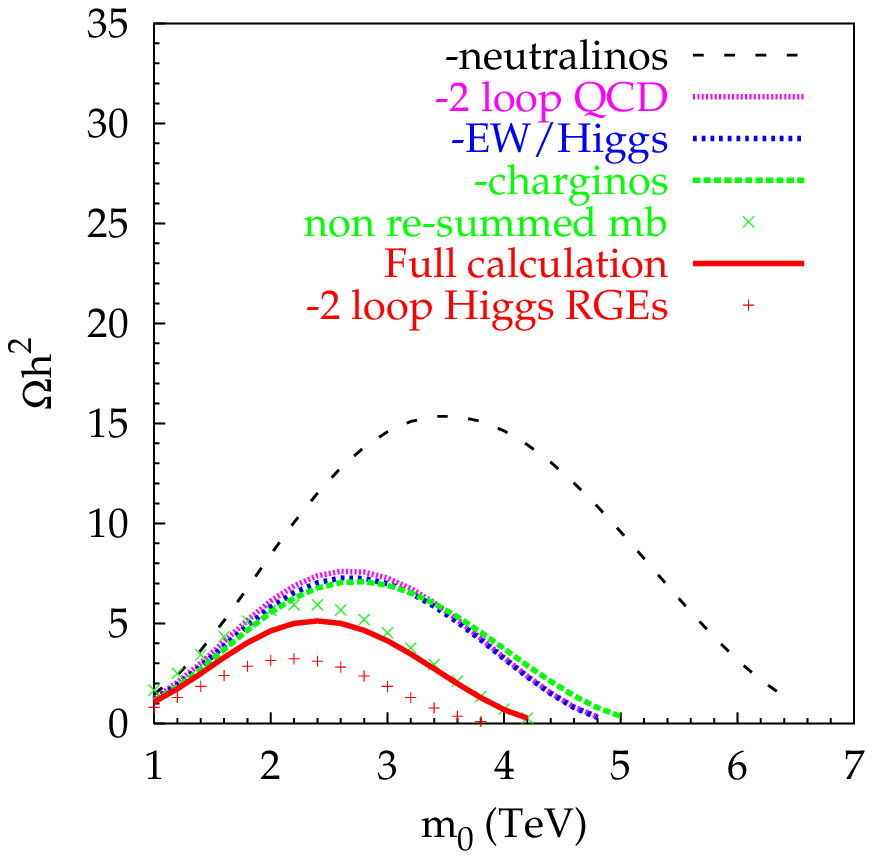}
\caption{(a)$\Omega h^2$ along slope S3. The central line
  gives the default {\tt SOFTSUSY} prediction, with the broken lines
  representing the limits due to scale uncertainty.
 The range of the abscissa corresponds to
  $M_{1/2}=440-1000$GeV. (b)Relic density dependence upon different approximations in
{\tt SOFTSUSY} for $\mhf=495$GeV, $A_0=0$, $\tan \beta=50$ and
$\mu>0$. The curves refer to the following approximations: (full
calculation): the default full calculation,
    (2 loop QCD): the default calculation with the 2-loop QCD
    corrections removed from the extraction of ${m_t}^{\overline{DR}}_{MSSM}
    (m_t)$, (EW/Higgs): the default calculation with the one-loop electroweak and
    Higgs corrections removed from the extraction of ${m_t}^{\overline{DR}}_{MSSM}
    (m_t)$, (neutralinos): the default calculation with the one-loop neutralino
    corrections removed from the extraction of ${m_t}^{\overline{DR}}_{MSSM}
    (m_t)$, (charginos): the default calculation with the one-loop chargino
    corrections removed from the extraction of ${m_t}^{\overline{DR}}_{MSSM}
    (m_t)$, (2-loops Higgs RGE):  the default calculation with only 1-loop RGEs for the Higgs
    potential soft SUSY     breaking parameters $m_{H_1}^2$, $m_{H_2}^2$,
    $m_3^2$, the 2-loops Higgs RGE are thus removed. (non-resummed $m_b$): the default calculation
    with no re-summation in SUSY $m_b$ corrections. \label{fig:foc}
%\label{fig:FPapp}
}}

It can be seen from Fig.~\ref{fig:foc}(a) that the prediction of
the relic density for the default {\tt SOFTSUSY}~prediction (the
unbroken line) is consistent with the WMAP measurement along S3
between $\mhf=440-1000$ GeV (corresponding to $m_0=4185-5611$
GeV). However, when $M_{SUSY}$ is varied by a factor of 2 in each
direction, the prediction can increase by up to 20$\%$ and
decrease by up to 30$\%$. This illustrates that the theoretical
uncertainties are large for the focus point regime.

 We now examine the theoretical uncertainty on the
predictions in more detail. We use various approximations when
calculating the sparticle mass spectrum and plot different values
predicted for $\Omega h^2$ as a function of $m_0$, for
$\mhf=495$GeV, $A_0=0$, $\tan \beta=50$, $\mu>0$. We are therefore
moving away from slope S3 to show that different approximations do
lead to different parameterisations. The results are displayed in
Fig.~\ref{fig:foc}(b). We can see the extreme sensitivity to loop
effects from the vastly different values of $m_0$ that agree with
the WMAP constraints from the figure. The Higgs mass parameter RG
evolution becomes very steep near $M_{SUSY}$, and so the
sensitivity to the number of loops used to evolve is high. The
high sensitivity to the extracted value of $h_t$ is shown by the
high sensitivity to the approximations in the calculation of
${m_t}^{\overline{DR}}_{MSSM} (m_t)$. At high\footnote{For the
$\tan \beta=10$ slope, there is no visible difference between the
calculation using re-summed or non-re-summed SUSY $m_b$
corrections.} $\tan \beta=50$, $h_b$, the bottom Yukawa coupling
is large and affects the running of $m_{H_1}^2$ appreciably. Its
extraction, as we have seen in the previous section, requires
knowledge of $\Delta \mb$. There was no visible difference from
the default calculation by using only one loop running for the
gaugino masses $M_{1,2,3}$ instead of the default two-loop
running, or removing all 2-loop terms while calculating the Higgs
masses and electroweak symmetry breaking conditions.

\subsection{Accuracies in mSUGRA}
\FIGURE{
 \unitlength=1.1in
 \begin{picture}(5.8,2.6)
 \put(-0.1,0){\epsfig{file=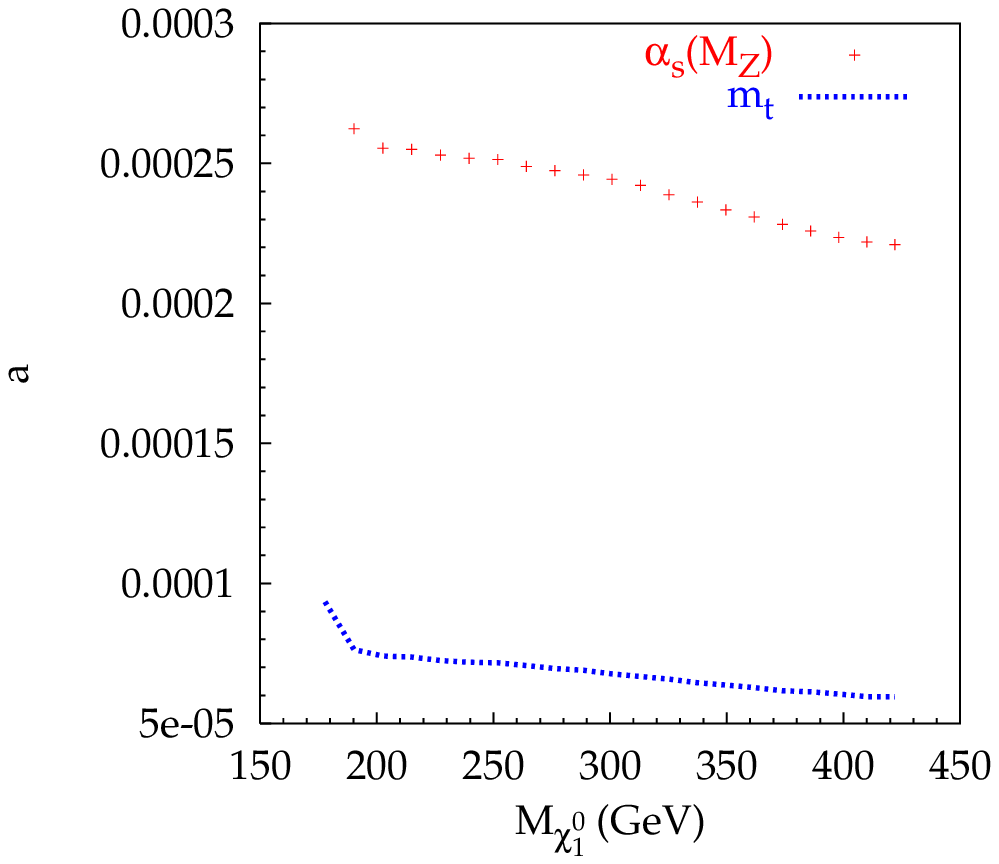, width=\wth}}
 \put(2.7,0){\epsfig{file=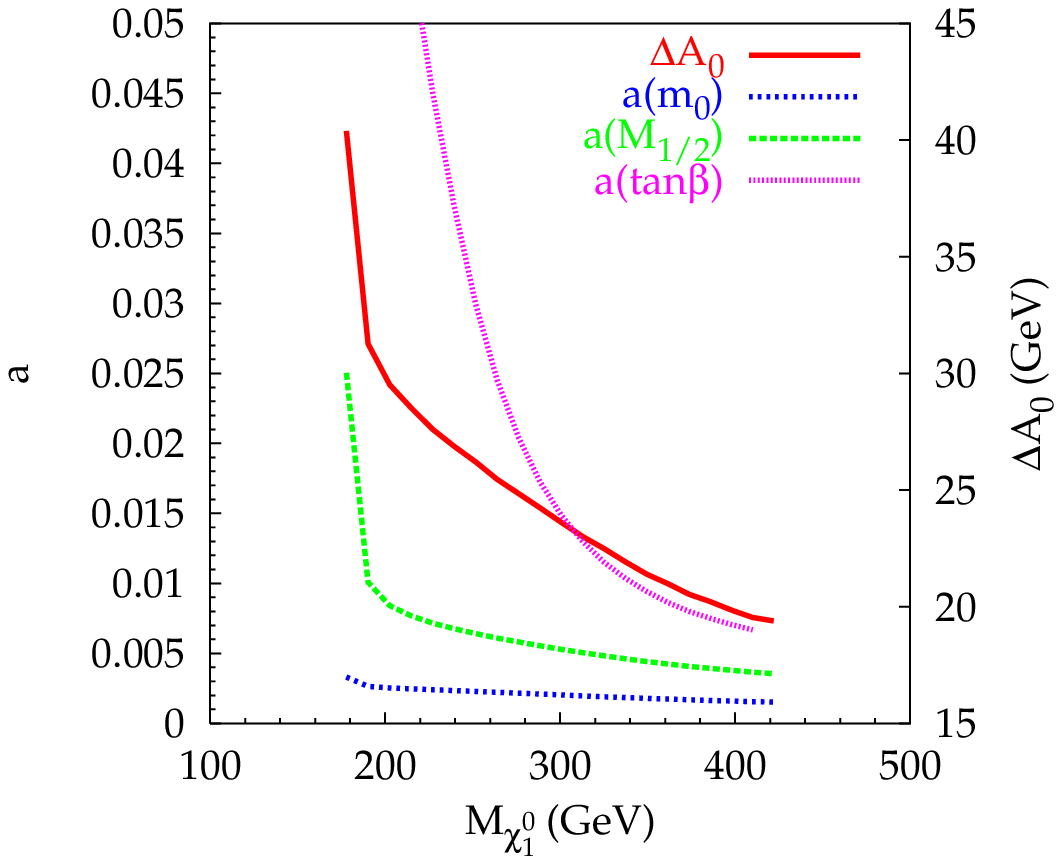, width=\wth}}
 \put(1.5,-.22){(a)}
 \put(1.3,0.5){\epsfig{file=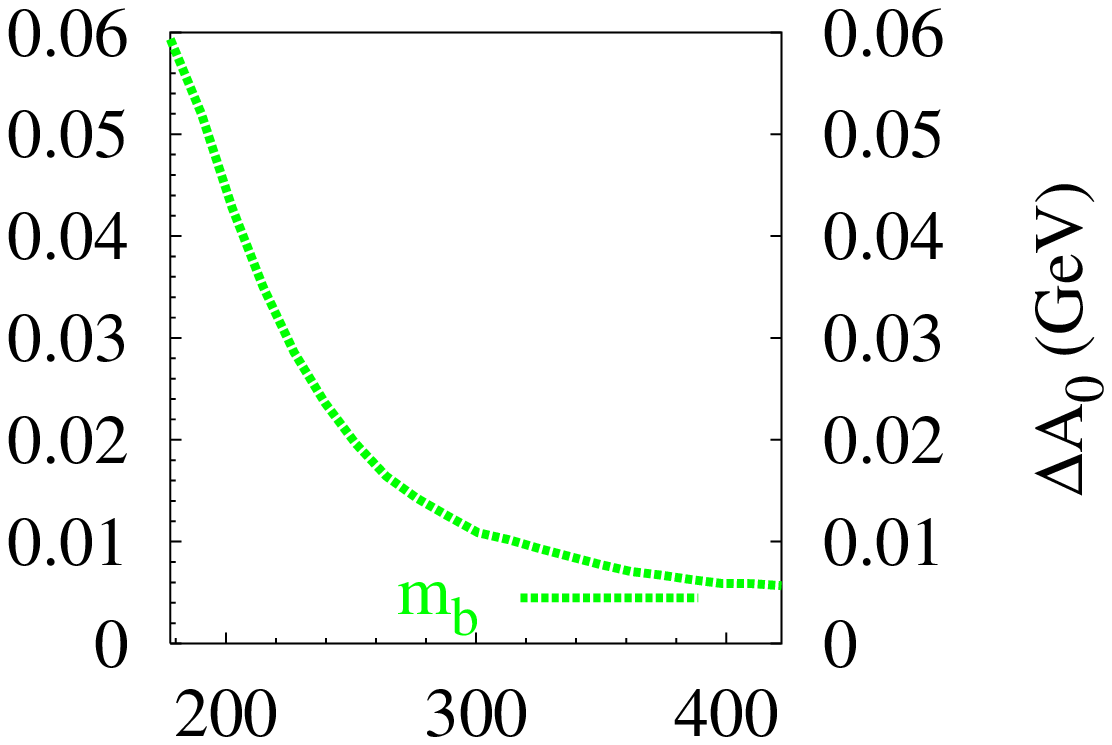, width=0.5 \wth}}
 \put(4.3,-0.22){(b)}
 \end{picture}
%\twographs{focusThDep.eps}{focusExpDep.eps}
\caption{Fractional accuracy required to achieve WMAP precision in
the mSUGRA
  scenario along slope   S3: (a) Standard Model inputs,
  (b) mSUGRA inputs.
 The range of the abscissa corresponds to
  $M_{1/2}=440-1000$ GeV.
  \label{fig:focdep}}}

We now turn to the issue of sensitivity of the $\Omega h^2$
prediction to the inputs in the focus point region. As explained
in section 2, we imagine that fits to LHC data have identified the
mSUGRA SUSY breaking parameters.  This would then allow one to derive
the entire set of the weak scale parameters, in particular all of
those which are of relevance for the calculation of the relic
density. As usual it is not only the mSUGRA SUSY parameters at
high scales which are needed but also the SM input parameters.
Considering what we have just seen concerning the crucial role
played by the top Yukawa coupling both in the RGE and in the
$t\bar t$ annihilation, we expect that a very precise input top
pole mass is required as well as $\alpha_s(M_Z)$, one of the
parameters that strongly affects the extraction of
${h_t}^{\overline{DR}}_{MSSM}$ from $m_t$. Fig.~\ref{fig:focdep}a
shows that there is indeed a huge sensitivity to $m_t$ and even
$\alpha_s(M_Z)$. $a(m_b(m_b))
\sim
\mathcal{O}(1-15)\%$ is shown as an insert in the
figure\footnote{For a
  $\tan \beta=10$, $m_t=172$ GeV focus point slope,
  $a(m_b(m_b))>1.0$.}.
Values of $a(m_t) \sim 10^{-4}$ mean that empirical uncertainties
of around 20 MeV on $m_t$ would be required for WMAP precision.
Studies suggest that $\delta(m_t)
\sim 30$ MeV is possible for the statistical
error~\cite{Aguilar-Saavedra:2001rg} at a suitable linear
collider, but that theoretical errors of order $100$MeV are
likely. $\alpha_s(M_Z)^{\overline{MS}}_{SM}$ must be measured to
better than an order of magnitude more precisely than current data
allow.

Fig.~\ref{fig:focdep}b shows that $m_0$ must be measured to a
fractional precision of better than $0.2\%$ for WMAP accuracy.
Such an accuracy is inconceivable at the LHC because, among other
reasons, the production cross-sections for the multi-TeV scalars
present in the focus-point region are tiny. In mSUGRA, $m_0$
determines (among others things) the value of $\mu$ which
determines the amount of Higgsino content.
%Since $m_0$ is very
%large in our scenario, a small change in $\mu$ ($m_0$) amounts to
%an extremely small relative change.
$M_{1/2}$ also affects the determination of $\mu$ as well as  the
mass of the LSP. We find that $\mhf$ should be measured to better
than $0.5\%$ for WMAP accuracy, and $A_0$ to better than 180 GeV,
provided $m_t$ is not quasi-degenerate with $M_{\neuto}$. The
sudden deterioration in the accuracy for $M_{\neuto} \sim m_t$ is
due to the approach to the $t$ threshold.  A similar behaviour is
seen in the required accuracy of $\tgb$. The measurement of $\tgb$
in  mSUGRA will be very demanding, due not only to the importance
of the $t \bar t$ vertex but also the extraction of the top Yukawa
that enters the RGE.

\subsection{Accuracies in PmSUGRA}
\FIGURE{ \epsfig{file=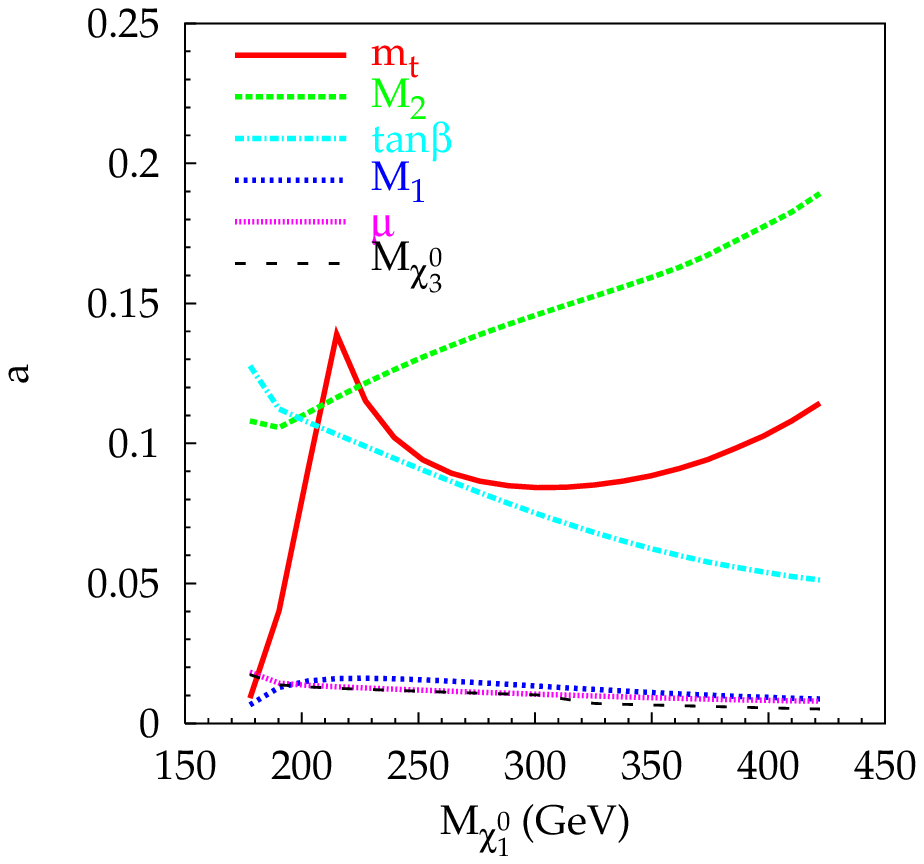,width=1.5\wth}
\caption{Required fractional accuracy upon MSSM parameters in the
PmSUGRA scenario along   slope S3. The abscissas correspond to
$\mhf=440-1000$ GeV.}\label{fig:focLC} }

Fig.~\ref{fig:focLC} displays the required accuracies $a$ for
several parameters in the PmSUGRA scenario. The relevant
parameters are the weak scale values of the elements of the
neutralino mass matrix. In particular, $M_1$ and $\mu$ are
important since, as discussed already, they determine the Higgsino
component of the LSP and how large the couplings of the LSP to the
Higgs bosons, $W$ and $Z$ are. $\tgb$ enters in the contribution
of the $b \bar b$ annihilation and in the neutralino couplings.
Since $M_2$ is not completely decoupled its effect is expected to
feed in as well. All of these parameters are essential for the
co-annihilation processes that set in for higher LSP masses.
Finally, there is also dependence upon $M_A$. Before we discuss
these accuracies and in order to highlight the contrast with the
mSUGRA approach, we comment on the accuracy needed on $m_t$. $m_t$
enters through the coupling of the neutral Goldstone boson as
$m_t^2$ (taking the dominant contribution) when the top threshold
sets in. We find a sensitivity to $m_t$ which is typically of the
same order as the precision on the relic density.  The present
empirical precision on $m_t$ is sufficient even for the PLANCK
benchmark. At the top threshold  one requires a much better
accuracy. On the other hand, the accuracy is less demanding for
higher LSP masses since the contribution of the $t\bar t$
annihilation is smaller. A rough estimate on $a(m_t)$, away from
threshold, is $a(m_t)\sim 1/(2c_t)$ in units of the precision on
the relic and where $c_t$ is the relative contribution of the $t
\bar t$ channel. $c_t$ can be read off from
Fig.~\ref{fig:focus-contrib}. Similarly in the whole range of
$M_{1/2}$ considered, $a(m_b)>3.0$ meaning that even the present
empitical $m_b$ errors reach PLANCK accuracy. The sensitivities of
$\mu,M_1$ require fractional precisions of 1$\%$ for WMAP
accuracy. This rather demanding accuracy originates more from the
couplings of the LSP to the Goldstone and $A$ pseudo-scalar
bosons. As we show in the figure, the accuracies on $M_1$ and
$\mu$ can be converted into accuracies on $\mneuto$ and
$\mneutth$. The accuracy needed on $M_2$ is  an order of magnitude
worse, though it is still non trivial to obtain empirically.
Therefore we see that if one can reconstruct the neutralino mass
matrix, which we think can be done rather precisely at the LC
provided there is enough energy to produce all of the neutralinos,
this scenario can be very much constrained. It could be
interesting to study whether this could be done without direct
production of the heaviest (wino-like) neutralino and how a
combination of LHC and LC data could be performed. A slight
problem still remains as concerns the $M_A$ dependence and its
required accuracy. In this scenario $M_A$ is very large, in excess
of $1$TeV and could be difficult to access at the LHC or a sub-TeV
LC. We find that at the lower end of the focus region, a change of
$50\%$ or  {\em higher} in $M_A$ only amounts to a maximum $5\%$
change in the relic density. However, at the higher end, though
$M_A\sim 1.5TeV$, a $30\%$ change in $M_A$ affects the relic
density by $10\%$. The reason that a larger (positive) change does
not deteriorate the precision is due to  the propagator
suppression of $M_A$ which means that the $A$ exchange no longer
contributes for too high values of $M_A$.

\section {Summary and Conclusions \label{sec:conc}}

Assuming mSUGRA,
we have examined the requirements for measurements of the MSSM spectrum in
order to
make reliable predictions of the relic density $\Omega h^2$. If such
requirements could be met, the assumptions that go into the derivation of
$\Omega h^2$ may be tested.

We find that theoretical uncertainties in the
mSUGRA prediction of $\Omega h^2$ {\em assuming standard cosmology} are
too high compared with the WMAP prediction.
In the co-annihilation scenario, the theoretical uncertainty
in $\Omega h^2$ (estimated by scale dependence)
is of the same size to twice as big as the WMAP accuracy. The situation is
worse in the problematic focus point and higher mass Higgs funnel regions,
where the  theory scale variation
is around five times bigger than the WMAP accuracy. In the lighter Higgs
funnel region, the scale variation is smaller than the WMAP accuracy.
The theoretical uncertainties are not a serious problem,
since
higher order calculations should be possible given enough time, effort and
motivation. If SUSY particles are discovered at a future collider, it seems
highly likely that this task will be accomplished.

The experimental information necessary in order to achieve a
precise prediction of the relic density is demanding. If one
assumes mSUGRA, fits must constrain $M_{1/2}$ and $m_0$ at the one
or two percent level and $A_0$ to roughly 100 GeV, for the
co-annihilation region for example. It is estimated in a
bulk-region test case~\cite{Battaglia:2003ab} that at the LHC this
accuracy is not quite achievable, even when the parameters are
quite favourable (for example $m_0=103 \pm 8$ GeV,
$M_{1/2}=240.0\pm3$ GeV at post-WMAP benchmark point B), but is an
order of magnitude too large at other points ($m_0=400\pm100$ GeV
and $M_{1/2}=400\pm8$ GeV at LHC points
1,2~\cite{Armstrong:1994it}). Combining complementary information
from the LHC and a future linear collider should provide the
necessary precision,
however~\cite{Blair:2000gy,Blair:2002pg,Allanach:2004ud}. Not only
must one experimentally control the soft SUSY breaking parameters
assuming mSUGRA, but Standard Model inputs must also be controlled
if the calculation of the MSSM spectrum is to be trusted. In the
co-annihilation region this is not a big problem, since the  parts
of the spectrum relevant to the prediction of $\Omega h^2$ (the
lightest neutralino and stau) do not have a large sensitivity to
the Standard Model inputs. The sensitivity in the focus point
region to the top Yukawa coupling $h_t$ is so high that $m_t$ must
be measured to better than 20 MeV. This is smaller than the strong
scale of the QCD interaction and therefore cannot be
perturbatively defined with such precision, rendering the approach
practically hopeless. In the Higgs funnel region, $m_t$ must be
measured to around 200 MeV, which may be achieved at a future
linear collider facility, but $m_b(m_b)$ must be constrained to
the 2-10 per mille level because of its effect on the mass of the
pseudo-scalar Higgs, which participates in the $s$-channel to
neutralino annihilation. This will not be possible since it is
much smaller than non-perturbative effects, but by measuring the
mass of the $A$ at the percent level, the sensitive dependence
upon $m_b(m_b)$ can be removed. For the heavier decoupled values
of $m_A$, a linear collider with sufficient centre of mass energy
will be required since the LHC will not provide enough precision
on measurements of the $A$ bosons~\cite{marco}.

Rather than relying upon mSUGRA to make predictions for the
spectrum, a more general strategy is to first determine what
regime one is in. If quasi-degenerate lightest stau and
neutralinos are discovered for example, it would indicate a
co-annihilation regime. In that case the mass splitting (less than
12 GeV) must be measured to better than $1$GeV in order to achieve
WMAP precision upon $\Omega h^2$ for LSP and $\stauo$ weighing
between $150$ to $450$GeV. In the focus point regime (indicated by
the measurement of gauginos but non-observation of scalars at the
LHC), the extreme sensitivity to $m_t$ vastly reduces if one no
longer has to assume mSUGRA and relies on individual mass
measurements instead. However, a measurement of $\mu$ and $M_1$ to
1$\%$ accuracy is required and will require copious production of
neutralinos and charginos at a linear collider with sufficient
centre of mass energy~\cite{Choi:2001ww,Choi:2000ta} since the
masses involved are high. In the Higgs funnel scenario, the
quantity $2 M_{\neuto} - M_A$ must be measured very precisely
requiring a measurement of $M_A$ and $\mneuto$ with a precision
ranging from $2\%$ for LSP masses around $100$GeV to $0.2\%$ for
LSP masses around $450$GeV.

The analysis we have performed in this paper was done in the
mSUGRA framework. WMAP now allows only three scenarios in mSUGRA
which are quite contrived and are challenging both from the
theoretical and experimental point of view. Going beyond mSUGRA,
for example in a scenario with non universal gaugino masses and in
particular a wino-like LSP, would be subject to less theory
uncertainty and might call for accuracies that are less demanding
for the colliders. It is also important to stress that we have
assumed that the standard derivation of the relic density is
correct. However, one should not forget that most alternatives
that affect the cosmological assumptions usually predict orders of
magnitude differences with the orthodox approach, far beyond the
few percent precision we have taken as a benchmark. This means
that perhaps some of these alternatives could be dismissed without
requiring too much precision. On the other hand it would be
interesting, assuming such schemes to be correct, to inquire about
the level of accuracy on the particle physics parameters that the
colliders must meet to corroborate some of these non-standard
cosmologies.

\section*{Acknowledgements}
We would like to thank M. Battaglia, P. Gondolo, J. Lesgourges, M.
Peskin, F. Richard, P. Salati and the LC-cosmology working group
for helpful conversations. This work was supported in part  by
GDRI-ACPP of CNRS and by  a grant from the Russian Federal Agency
for Science, NS-1685.2003.2.

\appendix

\section{Inputs\label{sec:sminp}}
Except where explicitly mentioned, we use the following default Standard Model
parameters as inputs~\cite{PDBook}:
\begin{eqnarray}
\alpha_s(M_Z)^{\overline{MS}}_{SM} &=& 0.1172,\
\alpha (M_Z)^{\overline{MS}}_{SM} = 1 / (127.934),\
m_b(m_b)^{\overline{MS}} = 4.23\mbox{~GeV}, \nonumber \\
m_{\tilde \tau} &=& 1.7777\mbox{~GeV},\
M_Z = 91.1876~\mbox{~GeV},\
m_t = 175\mbox{~GeV}.
\end{eqnarray}
A subscript $SM$ denotes the fact that the quantity has been derived assuming
Standard Model field content. $m_b(m_b)^{\overline{MS}}$ is derived from
experiment in 5-flavour QCD. A $\overline{MS}$ superscript denotes a quantity
calculated in the modified minimal subscription renormalisation scheme.
The weak mixing angle is derived from the muon decay constant $G_\mu = 1.16637
\times 10^{-5}$ GeV$^{-2}$.

The mSUGRA parameters are specified as $\tan \beta$, the ratio of the two
Higgs vacuum expectation values, the universal gaugino mass $M_{1/2}$, the
universal trilinear scalar coupling $A_0$ and the
universal scalar mass $m_0$. Universality is imposed upon the soft SUSY
breaking terms at a scale $M_{GUT}$ which is defined by
\begin{equation}
g_1(M_{GUT}) = g_2(M_{GUT}),
\end{equation}
$g_1$ and $g_2$ being the electroweak gauge couplings in the modified
$\overline{DR}$ scheme. $g_1$ is related to the Standard Model hypercharge
coupling $g'$ by the GUT normalisation
$g_1= \sqrt{5/3} g'$.
$g_3$ is set by the $\alpha_s(M_Z)^{\overline{MS}}_{SM}$ input and is not
unified with $g_{1,2}$.
We will take the Higgs potential parameter $\mu>0$, with its magnitude being
fixed by the electroweak symmetry breaking conditions~\cite{Allanach:2001kg}.

%\bibliography{dark}
\providecommand{\href}[2]{#2}\begingroup\raggedright\endgroup

\end{document}